\documentclass[aps,prc,showpacs,amsmath,amssymb]{revtex4}
\usepackage{graphicx}
\usepackage{dcolumn}
\usepackage{amsmath}
\renewcommand{\vec}[1]{\boldsymbol{#1}}

\begin{document} 

{\sf

\title{\Large Chiral vibrations in the A=135 region}  
\author{Daniel Almehed} 
\email{almehed@gmail.com}  
\affiliation{Department of Physics, University of Notre Dame,  
  Notre Dame, IN 46556, USA} 
\author{Friedrich D\"onau}
\email{doenau@hzdr.de}
\affiliation{Institut f\"ur Strahlenphysik, Helmholtz-Zentrum 
Dresden-Rossendorf, 01314 Dresden, Germany}    
\author{Stefan Frauendorf}  
\email{sfrauend@nd.edu} 
\affiliation{Department of Physics, University of Notre Dame,  
  Notre Dame, IN 46556, USA} 
\affiliation{Institut f\"ur Strahlenphysik, Helmholtz-Zentrum 
Dresden-Rossendorf, 01314 Dresden, Germany}   
\date{\today} 
\begin{abstract} 
Chiral vibrations in the A=135 region are studied in the framework of a 
RPA plus self-consistent tilted axis cranking formalism. 
In this model 
chiral vibrations appear as a precursor towards the static chiral regime. 
The properties of the RPA phonons are discussed and compared to  
experimental data. We discuss the limits the chiral region and the 
transition to the non harmonic regime. 
\end{abstract}  
\pacs{21.60.-n, 21.60.Jz, 21.30.-x} 
\maketitle  
 
\section{Introduction} 
Chirality is an important symmetry in many physical systems. It is a static property of the geometry of 
molecules with more than four different atoms, which can have  left-handed  
and  right-handed enantiomers. In these cases the left- and right-handed geometry  
is connected by reflection in a plane. In particle physics chirality 
is a dynamical feature of mass-less particles, which indicates  
 orientation of the intrinsic spin relative  
to the linear momentum. In a rotating nucleus it arises as a combination of 
a dynamical property, the angular momentum, with a static property,  
 the reflection symmetric triaxial shape of the  nucleus.  
If the  angular momentum vector lays outside any of the three principal planes, there  
are a left-handed and a right-handed arrangement,  
which are connected by the time reversal operation.  
This kind of chirality is  
manifested as a pair of (almost) degenerate  
rotational bands with the same parity~\cite{Fr01}.  
Pairs of such bands have been seen experimentally in the mass regions A=105, A=135 
and A=190 and theoretically described using mean field tilted axis  
cranking (TAC) models~\cite{DF00,OD04}, two-particle-rotor  
models~\cite{FM97,SC02} or extensions of the IBA model~\cite{BV04,TA06}. 
 
Chirality appears both in molecules and nuclei as a spontaneously broken 
symmetry: The many-body Hamiltonian describing these systems is invariant  
with respect to the chiral operation (space inversion or time reversal, respectively),  
which has the  consequence that the exact eigenfunctions are achiral.  
However, there exist very good approximate solutions, which are chiral, where 
the left-handed and right-handed configurations have the same energy.  
The exact eigenfunctions are odd and even superpositions of these  
chiral solutions. The energy splitting between these states is given 
by the interaction matrix element between the two chiral configurations, which is 
the inverse time for tunneling from one to the other. In most molecules 
the tunneling time is so long that they stay in one of the enantiomers, 
and the level splitting is unmeasurable. However, for  
CH$_3$NHF the tunneling is rapid. The 
splitting is in the order of 100 meV, i.e. the tunneling frequency is 
in the order of 3000 GHz, which is somewhat larger than the average  
rotational frequency of the molecule at room temperature.   
For the known nuclear cases, the left-right conversion 
is always rapid. The observed energy distance between the chiral partners is 
in the order of 100 keV~\cite{SK01}, except in very narrow spin regions, where  
the bands cross. This is comparable with the rotational frequency in the observed  
spin range.  
 
TAC is a microscopic mean field method that has been shown   
to very well describe the energy and the intra-band transition rates of  
the lower of the two chiral partner bands, see e.g.~\cite{HB01}. 
 However, it gives either 
 one achiral self-consistent solution or two degenerate chiral ones. 
It can not describe the left-right mode, which is a well known deficiency  
of the mean field solution when it breaks spontaneously a symmetry 
(cf. e.g.~\cite{RS80,Fr01}).  
In a finite system, as the nucleus, the symmetry breaking develops in a gradual way. 
First a precursor appears as a slow vibration around the symmetric configuration. 
 Becoming increasingly anharmonic, the vibration changes into tunneling between 
two asymmetric configurations, which is progressively inhibited. 
 
In order to describe the splitting between the two bands 
one has to go beyond the mean field approximation. So far this has only 
be done in the framework of  two-particle-core coupling models~\cite{FM97,SC02,BV04,TA06}.  
These studies show a development from chiral vibrations toward 
tunneling between static left- and right-handed configurations with increasing  
angular momentum.  Although the quantal nature allow the core-particle models to 
account well for this aspect of symmetry breaking,  
they are based on several assumptions the validity  
of which is not assured. Examples are: Rigid shape and irrotational-flow moments of inertia 
for a triaxial rotor core~\cite{FM97,SC02}, which are not consistent  
with microscopic cranking calculations~\cite{TA06}, 
the application of the IBA core, and the IBAFF  
coupling scheme~\cite{BV04,TA06}. A microscopic treatment starting from the 
TAC mean field seems important for a better understanding of nuclear chirality.  
Obviously the transition from chiral vibrations to chiral rotation involves  
large amplitude collective motion, a treatment of which goes beyond the scope  
of this paper. However the regime of  chiral vibration can be treated by the  
Random Phase Approximation  (RPA) approach as long as the energy splitting  
between the bands is large enough for the assumption of a  harmonic vibration  
remaining reasonable. As the energy splitting decreases the anharmonic effects  
will become more important until the transition point where the RPA energy  
becomes zero and the TAC mean field becomes chiral. The present paper is devoted  
to  a study of the spin range before the transition point where the RPA  
approximation will allow us investigating the chiral vibrations and provide  
insight into the nature of nuclear chirality.  
 
In the static chiral picture one would expect the two bands to have very similar 
electromagnetic transition rates. Recent experiments have shown different intra-band  
transition rates in the two chiral partner bands in $^{134}$Pr~\cite{TA06,PH06} while  
the bands in $^{128}$Cs~\cite{GS06} and $^{135}$Nd~\cite{MA07} have similar transition  
rates. The different transition rates in $^{134}$Pr have been interpreted  
as being due to coupling of shape degrees of freedom to the orientation degree of  
freedom~\cite{TA06} or even that the bands are not chiral partner bands at  
all~\cite{PH06}. 
 
Chiral pairs of rotational bands  have mainly been suggested for nuclei with  
odd proton and odd neutron number. 
In the A=135 region, they are build on the  
$\pi h_{11/2} \otimes \nu h_{11/2}^{-1}$ configuration. Candidate bands  
have been seen in the odd-odd nuclei $^{126-132}$Cs, $^{130-134}$La, $^{132-134}$Pr, 
$^{136}$Pm and $^{138-140}$Eu.   
In this work we focus on the N=75 isotone chain and the Z=57 isotope  
chain which represent the central part of this region. 
We have performed  calculations in the framework of tilted axis cranking  
(TAC) and random phase approximation (RPA) for the twin bands in the N=75  
isotones and the Z=57 isotopes built on the $\pi h_{11/2}\otimes \nu h_{11/2}^{-1}$ 
configuration.  
 Examples for chiral partner bands exist also in odd-even  
nuclei~\cite{ZG03,MA07}. The case of $^{135}$Nd has been studied in~\cite{MA07} by  
means of the method presented in this paper. In principle, chiral partner bands should also  
exist  in even-even nuclei build on two-particle two-hole like configurations.  
For these and 
 even more complex configurations our method can be directly applied.   
We describe the formalism in section~\ref{sec:Formalism}. The results for the  
energy, amplitudes and transition rates are presented in  
section~\ref{sec:Results}. An analysis of the structure of the phonons is given  
in section~\ref{sec:amplitude}, where it will be demonstrated that for most  
of the cases the chiral character prevails, i.e. the coupling 
to the shape degrees of freedom is weak. 
 
\section{Formalism} 
\label{sec:Formalism}

Most earlier TAC calculations for chiral bands have used the Strutinsky  
shell correction method for calculating energies and band properties 
(c.f. the SCTAC model ~\cite{Fr00}). In this paper we present selfconsistent 
RPA calculations which are founded on the TAC plus RPA with a residual 
quadrupole-quadrupole (QQ) interaction. The corresponding
Hamiltonian $\hat{H}$ which similar to the one of the PQTAC model \cite{Fr00} is given by
\begin{eqnarray} 
\label{eq:H1} 
\label{H1}
\hat{H}&=& 
\sum_{\tau =\pm 1}
[\,\,\hat{h}^\circ_{\tau}   
-\Delta_\tau (\hat{P}^\dagger_\tau + \hat{P}_\tau)  - \lambda_\tau \hat{N}_\tau  \,\,]
- \vec{\omega} \cdot \vec{\hat{J}} 
-\frac{\kappa_{_0}}{2} 
\sum_{m=-2,2} (-1)^{m}
\hat{Q}^{(BK)}_{m} \hat{Q}^{(BK)}_{-m}. 
\end{eqnarray} 
The first part $\hat{h}^\circ_{\tau}$ is the spherical single particle term of 
the Hybrid TAC model \cite{DF00} which on one hand takes advantage of using single particle 
energies of the spherical Woods-Saxon model and on second hand it uses the same 
oscillator basis with the spherical quantum numbers $\{n,l,j,m\}$  as applied here. 
The isospin index $\tau=\pm1$ distinguishes the neutron and proton contributions, respectively, 
 in the Hamiltonian. 
The second term $\Delta_\tau ( \hat{P}^\dagger_\tau + \hat{P}_\tau )$ means a pair field 
where $\hat{P}^\dagger_\tau$ and $\hat{P}_\tau $ are the familiar monopole pair operators.
The gap parameters $\Delta_{\tau}$ are 80\% of the odd-even mass differences 
to the respective neighboring nuclides.  
The third term $\lambda_\tau \hat{N}_\tau$  implying the particle
number operators $\hat{N}_\tau$ is introduced as usual for saving in average the  
numbers of protons and neutrons, Z and N, respectively by 
appropriate choice of the Fermi energy $\lambda_\tau$.
The forth term $-\,\vec{\omega} \cdot \vec{\hat{J}}$ in Eq.(\ref{eq:H1})
is a 3d-cranking  energy term. Therein the vector $\vec\omega=(\omega_1,\omega_2,\omega_3)$ 
defines the angular frequency $\omega=|\vec\omega|$ and the direction of the cranking axis 
(cf.~Eq.~(\ref{omega})). 
The operator $\vec{\hat{J}}$ means the total angular momentum operator. 

The last contribution in the Hamiltonian $\hat{H}$, Eq.~(\ref{eq:H1}) is an
isoscalar QQ interaction with the strength parameter $\kappa_{_0}$.
We adopt the modified quadrupole operators $\hat Q^{(BK)}_{m}$ 
of the collective model  by Baranger and Kumar (BK) \cite{BK68}  
\begin{eqnarray}
\label{Qop}
\hat Q^{(BK)}_{m}&=&\hat Q^{(BK)}_{m}(\tau =+1)+\hat Q^{(BK)}_{m}(\tau =-1)\\ 
\text{where}\quad\quad
  \hat Q^{(BK)}_{m}(\tau) &=& \left(\frac{n_{\rm low}-B}{\hat n-B}\right) 
  \left(\frac{2 A_{\tau}}{A}\right)^{1/3} r^2 Y_{2m} (\tau ),\quad\quad m=0,\pm 1,\pm 2\nonumber
\end{eqnarray} 
The operator $\hat n$ counts the number of oscillator quanta $n$, i.e.  $\hat n\,| nljm \rangle =n\,| nljm \rangle$.

Compared to the usual isoscalar quadrupole operator 
$\hat{Q}_{m}=r^2Y_{2m}(\tau =+1)+r^2Y_{2m}(\tau =-1)$ 
the operator $Q^{(KB)}_{m}$  in Eq.~(\ref{Qop}) contains additionally quenching factors
which depend on the isospin, the mass number $A$, the oscillator shell quantum number $n$  
and a stiffness parameter $B$. The oscillator basis of our calculations includes  all substates 
of the two oscillator shells $n=n_{low}=4$ and $n=5$.
The value of the parameter $B$ is chosen to approximately  reproducing potential energy 
surfaces obtained by the Strutinsky correction method~\cite{Fr00}. Setting the parameter 
$B=0.5$ gives a similar prolate-oblate mass difference and $\gamma$-softness as SCTAC 
calculations in  this mass region. In other mass regions a different value of $B$ 
may be needed. The strength parameters $\kappa_{_0}$ for the considered nuclides are listed in 
table~\ref{tab:kappa}. They are adjusted to fulfill the  self-consistency conditions 
(c.f. Eq. (\ref{scepsga})) at the minimum of the potential energy 
surfaces obtained by the Strutinsky shell correction method. This choice ensures that 
the two spurious RPA solutions appear at the predefined energies $E_{_{RPA}}=0$ and 
$E_{_{RPA}}=\hbar \omega$.  
In order to simplify the notation  we will leave out in what follows the suffix
$(BK)$ of the operator $Q^{(BK)}_{m}$ keeping in mind that we are working in 
this paper with the BK modification. 

\begin{table} 
\caption{\label{tab:kappa}Quadrupole force strength $\kappa_{_0}$.} 
\begin{ruledtabular} 
\begin{tabular}{c|cc} 
 & $\kappa_{_0}$  (MeV fm$^{-4}$) &\\ 
\hline 
$^{130}$Cs & 0.00354 & \\ 
$^{132}$La & 0.00343& \\ 
$^{134}$Pr & 0.00329 & \\ 
$^{136}$Pm & 0.00333 & \\ 
$^{138}$Eu & 0.00337 & \\ 
$^{140}$Tb & 0.00328& \\ 
$^{130}$La & 0.00359 & \\ 
$^{134}$La & 0.00319& \\ 
\end{tabular} 
\end{ruledtabular} 
\end{table} 
The first step of the calculation is the search for self-consistent (s.c.) TAC solutions,  
which for a fixed strength $\kappa_{_0}$ and within the considered frequency range 
$\omega$ = 0.1 - 0.4 MeV   
correspond to a {\it tilted} cranking solution. The general scheme of these 
calculations has been described in Ref.~\cite{Fr00}. 
 The deformed potential $v$ produced by 
the mean field contribution from the QQ interaction has the form
\begin{equation}
\label{Qmf}
         v=v(q_{_0},q_{_2}) =\kappa_{_0} q_{_0} \hat Q_{_0}+ \kappa_{_0}q_{_2}(\hat Q_{_2}+\hat Q_{_{-2}}),
\end{equation}
where $q_{_0}$ and $q_{_2}$ are defined by the expectation values 
\begin{eqnarray}
\label{sc}
\langle \hat Q_{_0}\rangle&=&q_{_0}\nonumber\\
\langle \hat Q_{_2}\rangle&=&\langle \hat Q_{_{-2}}\rangle =q_{_2},
\end{eqnarray}
and $\rangle$ is the short hand notation for the TAC solution  $|TAC\rangle$ for the 
selected configuration. 
The absence of the components $\hat Q_{_{\pm1}}$ in the deformed potential (\ref{Qmf}) ensures 
that the condition $\langle \hat Q_{_{1}}\rangle=\langle \hat Q_{_{-1}}\rangle=0$ is satisfied
and, therefore,  one stays  in the principal axes (P-) system 
of the uniformly rotating ellipsoidal potential. 
This holds generally, i.e. also for mean field states with a tilted cranking axis. 
Eqs.(\ref{sc}) represent the two selfconsistency conditions for the nuclear shape.
Actually we are using the deformation parameters $(\varepsilon,\gamma)$ of  
the modified oscillator model~\cite{Nilsson-Ragnarsson}. Then the s.c. conditions take the form 
\begin{eqnarray}\label{scepsga}
\kappa_{_0}\langle \hat Q_{_0}\rangle=\kappa_{_0} q_{_0}(\varepsilon,\gamma) 
&=&\,\,\hbar\omega_{_0} \,2/3\,\varepsilon \cos{\gamma}\nonumber\\
\kappa_{_0}\langle \hat Q_{_2}\rangle=\kappa_{_0} q_{_2}(\varepsilon,\gamma) 
&=&-\hbar\omega_{_0} \,2/3\,\varepsilon \sin{\gamma}/\sqrt{2}.
\end{eqnarray}
where $\hbar\omega_{_0}=41A^{-1/3}$ is the usual oscillator frequency. 

The P-components of the rotational frequency vector 
$\vec{\omega}$ in Eq.(\ref{H1}) are expressed by the spherical angles 
$(\vartheta,\varphi)$: 
\begin{eqnarray}
\label{omega}
\omega_{_1}~&=&~\omega~\sin\vartheta\cos\varphi\nonumber\\ 
\omega_{_2}~&=&~\omega~\sin\vartheta\sin\varphi\\ 
\omega_{_3}~&=&~\omega~\cos\vartheta \nonumber~.
\end{eqnarray}
The SCTAC model includes the selfconsistency with respect to the spin orientation,
i.e. the stability of the angular momentum  vector $\vec J=\langle \vec {\hat J}\rangle$. The
s.c. orientation of the angular momentum $\vec J$ 
is determined by the parallel condition $ \vec\omega\Vert\vec J$ ~\cite{Fr00} which
fixes the angles $(\vartheta,\varphi)$ in Eq.(\ref{omega}) for given frequency $\omega$. Because 
the vector $\vec J$ is a constant of motion in the laboratory (L-) system, it is a
still-standing vector about which the deformed nucleus is uniformly rotating.  Hence we can chose  
the laboratory $z$-axis along this direction which fixes the transformation  
from the P- to the L-system needed below. 

In our TAC calculations for the A=135 region, chiral solutions appear   
as a transient phenomenon. With increasing frequency, the 
angular momentum vector moves   from the principal axis plane   
$\varphi=0$ spanned by the 1- (short) and 3- (long) axis   to  
the principal plane $|\varphi |=90^\circ$ spanned by the 2- (intermediate) and 3-(long) axis  
~\cite{DF00}.  This can be seen e.g. in Tab.\ref{tab:epsgam} for the case $^{136}$Pm.
The RPA calculations are based on the
regimes where the TAC angular momentum  lays {\it within} one of these principal planes. 
Then the excited chiral partner band can  be described in RPA as a chiral vibration, 
which is a periodic motion of the  orientation of the deformed potential  
relative to angular momentum vector $\vec J$, i.e. a vibration of the angles 
$\vartheta$ and   $\varphi$ about their equilibrium values. 
There are two critical frequencies. The first one is associated with the transition 
of the TAC mean field solution at $\varphi=0$ into 
the static chiral regime, and the second one is associated with the transition out  
of chiral regime at $|\varphi |=90^\circ$. These transitions  of the mean field are  
accompanied by instabilities of the RPA, where the energy of the lowest  
RPA phonon becomes zero thus indicating that the RPA is not applicable beyond the 
critical frequency. 
 
When the TAC solution has reached the static chiral regime $0<|\varphi|<90^\circ$ 
our present approach gives zero energy splitting between the two chiral  
configurations with $\varphi=\pm |\varphi|$. However, higher order terms 
in our Hamiltonian would give rise  
to tunneling between the left-handed and the right-handed  
solution which causes an energy splitting between the two bands in this 
region too. 
\begin{figure}[htbp] 
\centerline{\includegraphics[clip,width=8.5cm]{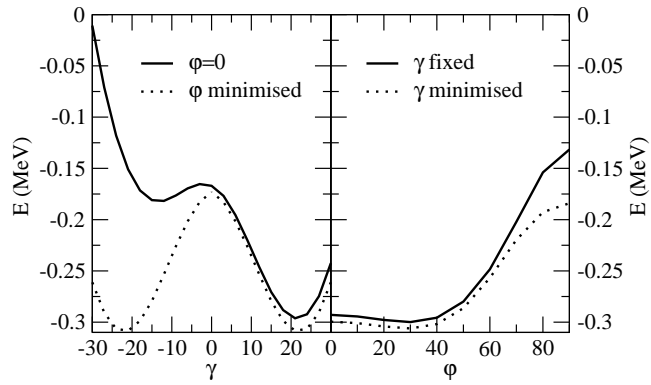}} 
  \caption{Potential energy in the rotating frame for $^{134}$Pr at  
    $\omega=0.4$ MeV/$\hbar$. 
    The two solid curves are calculated with $\gamma$ or $\varphi$ (in degrees) fixed while 
    the dotted curve is calculated with a full  minimization.} 
  \label{fig:pr134PESgf} 
\end{figure} 
Figure~\ref{fig:pr134PESgf} shows the TAC energy  in the rotating frame,  
spanned by 
the three principal axes of the triaxial density distribution, 
as a function of the tilt angle $\varphi$ and the triaxial deformation parameter  
$\gamma$. The calculation is done for the two quasi particle  (2qp) configuration  
$\pi h_{11/2} \otimes \nu h_{11/2}$ in $^{134}$Pr at $\omega=0.4$ MeV. 
The energy surface is rather shallow especially in the $\varphi$ direction. When we allow 
for the full minimization in the orientation degrees of freedom the surface becomes  
symmetric in $\pm \gamma$, because the change of sign is equivalent with a reorientation 
of the axes. It is therefore enough  to consider  $\gamma>0$ only. 
\begin{figure}[htbp] 
\centerline{\includegraphics[clip,width=8.5cm]{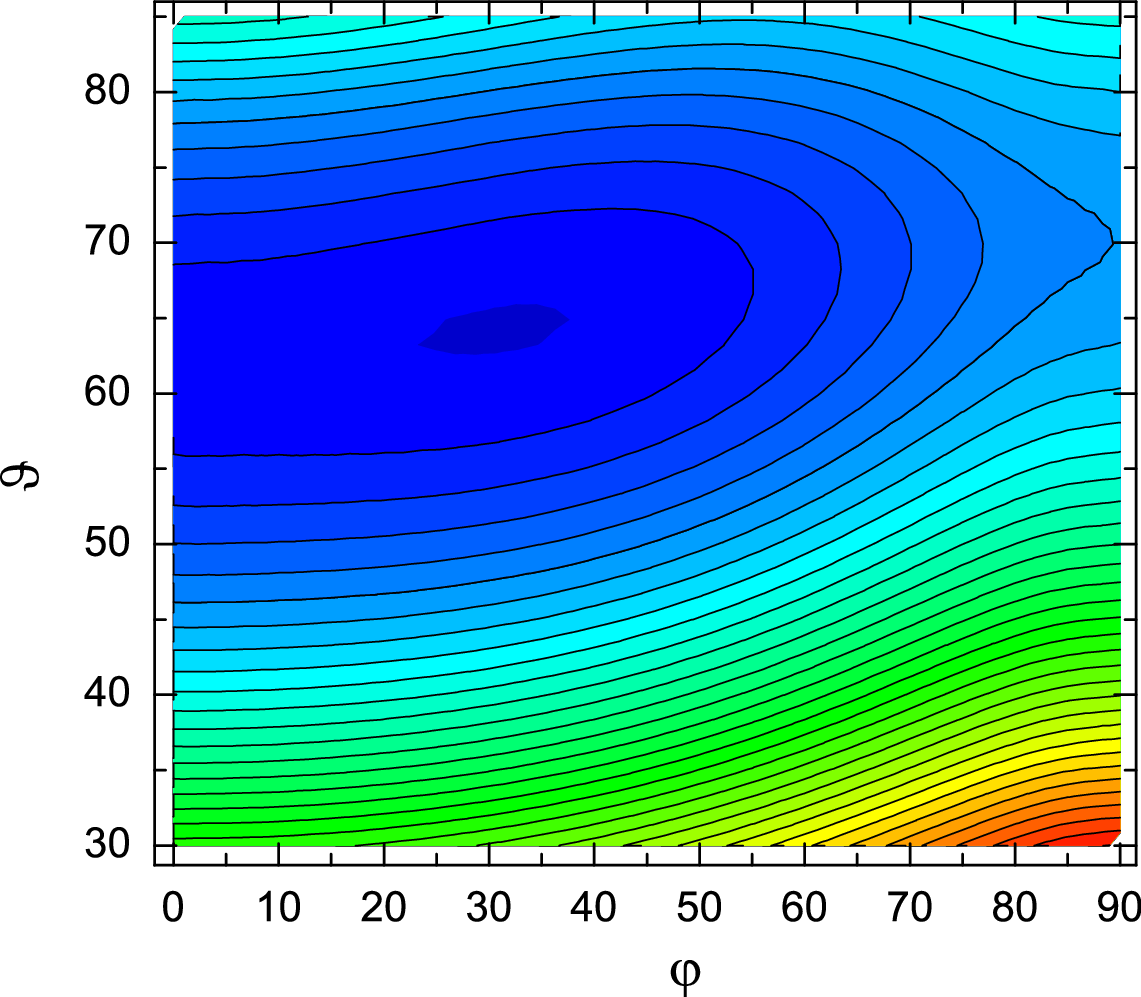}} 
  \caption{(Color online) Potential energy surface in the rotating frame for  
  the 2qp configuration $\pi h_{11/2} \otimes \nu h_{11/2}$ in 
    $^{134}$Pr at $\omega=0.4$ MeV/$\hbar$ as a function of the tilt angles 
$\vartheta$ and $\varphi$ (in degrees)   
    with constant deformation. The equipotential lines are separated by 25 keV. Darker  
    color represent lower energy.} 
  \label{fig:pr134PEStf} 
\end{figure} 

{\sf
\begin{table}[htbp]  
\caption{\label{tab:epsgam}
Equilibrium values  of the deformation parameters $\varepsilon$,$\gamma$, the tilt angles 
$\theta,\varphi$ and the lowest RPA phonon energy  $\hbar\Omega$  as a function of the 
cranking frequency $\omega$.
} 
\begin{ruledtabular} 
\begin{tabular}{ccccccc} 
Nuclid &  $\omega$(MeV/$\hbar$) & $\varepsilon$& $\gamma$(deg) &$\vartheta$(deg)&
$\varphi$(deg)&$\Omega$(MeV/$\hbar$)\\\hline\\
$^{130}$Cs
   &      0.20&        0.19&     33&          59&             0&            0.30\\
    &        0.25&        0.19&     33&          59&             0&            0.27\\
     &       0.30&        0.19&     32&          59&             0&            0.21\\
     &       0.35&        0.19&     31&          60&             0&            0.12\\
   &       0.40&        0.19&     31&          60&             19&             -\\
    &       0.45&        0.19&     29&          61&             34&            -\\\\
$^{132}$La 
  &    0.20&        0.21&     26&          59&             0&            0.33\\
   &         0.25&        0.21&     26&          60&             0&            0.30\\
   &         0.30&        0.21&     26&          61&             0&            0.26\\
   &        0.35&        0.21&     25&          62&             0&            0.19\\
   &         0.40&        0.21&     24&          63&             0&            0.09\\
   &         0.45&        0.21&     22&          64&             24 &           -\\\\
$^{134}$Pr
    &     0.10&        0.21&     24&          57&             0&            0.30\\
    &        0.15&        0.21&     24&          58&             0&            0.32\\           
    &        0.20&        0.21&     24&          59&             0&            0.31\\
   &         0.25&        0.21&     24&          60&             0&            0.28\\
   &         0.30&        0.21&     23&          61&             0&            0.21\\
    &        0.35&        0.21&     22&          61&             0&            0.11\\
    &        0.40&        0.21&     22&          64&             32&            -\\\\
$^{136}$Pm
   &      0.10&        0.24&     24&          50&             0&            0.14\\
    &      0.15&        0.24&     24&          50&             0&            0.09\\
    &        0.20&        0.24&     24&          55&             20&            -\\               
    &        0.25&        0.24&     24&          58&             38&            -\\               
    &        0.30&        0.24&     24&          62&             52&            -\\               
    &        0.35&        0.25&     24&          66&             77&              -\\               
    &        0.40&        0.25&     24&          69&             90&           0.15\\               
    &        0.45&        0.25&     21&          72&             90&           0.33\\               
     &       0.50&        0.25&     19&          75&             90&           0.42\\  \\
$^{138}$Eu
  &      0.05 &        0.30&     21&          39&             22&           -\\
  &      0.10 &        0.29&     21&          47&             90&           0.07\\
   &         0.15&        0.29&     21&          54&             90&           0.14\\         
    &        0.20&        0.29&     22&          58&             90&           0.20\\         
   &         0.25&        0.29&     22&          61&             90&           0.27\\         
   &         0.30&        0.29&     22&          63&             90&           0.34\\         
   &         0.35&        0.29&     22&          65&             90&           0.40\\         
   &         0.40&        0.29&     22&          67&             90&           0.46\\ \\
$^{140}$Tb
  &      0.20&        0.29&     23&          53&             90&           0.27\\         
   &         0.25&        0.29&     24&          57&             90&           0.35\\         
   &         0.30&        0.29&     25&          60&             90&           0.43\\         
   &         0.35&        0.28&     25&          64&             90&           0.50\\         
   &         0.40&        0.28&     26&          67&             90&           0.58\\ \\
$^{130}$La
     &      0.10&        0.26&      9&          61&              0&           0.38\\         
    &        0.15&        0.25&     12&          63&              0&           0.43\\         
    &        0.20&        0.25&     13&          65&              0&           0.48\\         
    &        0.25&        0.25&     11&          66&              0&           0.53\\         
    &        0.30&        0.25&      9&          68&              0&           0.59\\         
    &        0.35&        0.26&      8&          69&              0&           0.66\\ \\        
$^{134}$La 
    &      0.10&        0.15&     38&          51&              0&           0.32\\         
     &       0.15&        0.15&     37&          52&              0&           0.42\\         
     &       0.20&        0.15&     36&          52&              0&           0.43\\         
     &       0.25&        0.15&     35&          52&              0&           0.35\\         
     &       0.30&        0.15&     35&          52&              0&           0.27\\         
     &       0.35&        0.15&     34&          52&              0&           0.19\\         
     &        0.40&        0.15&     33&          53&              0&           0.10\\         
     &        0.45&        0.15&     33&          53&              55&           -\\         

\end{tabular} 
\end{ruledtabular} 
\end{table} 
 }

Figure~\ref{fig:pr134PEStf} shows the potential energy surface in the rotating frame 
(total routhian) as  a function of  both angles $\vartheta$ and $\varphi$. 
One can see the very soft nature of the potential in the $\varphi$ direction. 
 
The RPA accounts for the harmonic excitations above the mean field minimum. It 
gives an adequate description  as long as we are in the  chiral 
vibrational regime well before the transition to static chirality or after returning 
to the former regime.  
The transition point corresponds to the rotational frequency where the RPA excitation 
energy goes to zero. 
After solving the s.c. mean field problem the Hamiltonian (\ref{H1}) is expressed in 
terms of the quasiparticle creation (annihilation) operators 
$\hat{\alpha}_i^\dagger (\hat{\alpha}_i)$, and it takes the form
\begin{equation} 
  \hat{H} = \hat{h}_{\rm mf} + \hat{H}_{\rm res}, 
\end{equation} 
where $\hat{h}_{\rm mf}$ is the diagonal mean field Hamiltonian  
\begin{equation} 
  \hat{h}_{\rm mf} = E_{\rm mf} + \sum_i e_i \hat{\alpha}_i^\dagger \hat{\alpha}_i ,
\end{equation} 
 and  $\hat{H}_{\rm res}$ is the residual QQ interaction. 
We make use of the quasi-boson approximation 
$\hat{b}_\mu^\dagger = \hat{\alpha}_i^\dagger \hat{\alpha}_j^\dagger$, where 
the $\hat{b}_\mu^\dagger$    are treated as exact bosons, and introduce  the combined 
index $\mu  \equiv  \{ i>j, \tau \}$. 
The Hamiltonian $\hat{H}$ is rewritten in RPA order by only keeping  
terms up to second order in the boson operators~\cite{RS80}, i.e.
\begin{equation} 
  \hat{H}_{\rm RPA} = E_{\rm mf} + \sum_{\mu,\nu} A_{\mu\nu} \hat{b}_\mu^\dagger 
  \hat{b}_\nu + \frac{1}{2} \sum_{\mu,\nu} \left( B_{\mu\nu} \hat{b}_\mu^\dagger 
  \hat{b}_\nu^\dagger +{\rm h.c.} \right).
\end{equation} 
The matrices ${\bf A}$ and ${\bf B}$ are  real and symmetric,
\begin{eqnarray} 
  A_{\mu\nu} &=& \delta_{\mu\nu} E_\mu - 
\kappa_{_0}\sum_{m=-2}^2 \,q_\mu^m {q_\nu^m}^*,\nonumber\\ 
  B_{\mu\nu} &=& -\kappa_{_0}\sum_{m=-2}^2\, q_\mu^m {q_\nu^m}^*, 
\end{eqnarray} 
where $q_\mu^m=\langle\hat{b}_\mu \hat{Q}_m \rangle$ are the quadrupole matrix elements 
in quasiboson representation~\cite{RS80} and $E_\mu=e_i + e_j$ are the 2qp energies.  
The quadrupole operators $\hat{Q}_m$ take the form
\begin{eqnarray} 
  \hat{Q}_m &=& \sum_\mu  (q_\mu^m \hat{b}_\mu^\dagger + {q_\mu^m}^* \hat{b}_\mu).
\end{eqnarray} 
Note, the sum runs also  over $\tau=\pm1$. 
We solve the RPA equation 
\begin{equation} 
  \label{eq:rpa1} 
  \left[\hat{H}_{\rm RPA}, \;\,\hat{O}^\dagger_\lambda \right]= E_{_{RPA}}^\lambda  \hat{O}^\dagger_\lambda, 
\end{equation}  
using the strength function  
method of~\cite{RS80,KN86}. The RPA eigenmode operators $\hat{O}^\dagger_\lambda$ are 
\begin{equation} 
  \label{eq:rpa2} 
  \hat{O}^\dagger_\lambda = \sum_\mu (X^\lambda_\mu \hat{b}_\mu^\dagger - 
  Y^\lambda_\mu \hat{b}_\mu), 
\end{equation}  
where the RPA amplitudes $X^\lambda_\mu$ and $Y^\lambda_\mu$ are obtained 
 by solving the standard set of linear  
equations resulting from Eq.~(\ref{eq:rpa1}) together with the normalization condition 
\begin{equation} 
  \label{eq:rpanorm} 
  \left[ \hat{O}_\lambda , \;\,\hat{O}^\dagger_{\lambda'} \right] = 
  \sum_\mu (X^\lambda_\mu X^{\lambda'}_\mu -  
  Y^\lambda_\mu Y^{\lambda'}_\mu )= \delta_{\lambda \lambda'}. 
\end{equation} 
Since we use a separable force,  
this set of linear equations is strongly simplified~\cite{KN86}.  
 
As mentioned above there are two rotational spurious solutions in the RPA spectrum. One 
at zero energy $E_{_{RPA}}=0$ induced by the angular momentum operator in the L-frame, 
$\hat{J}_z$, and one at the rotational energy $E_{_{RPA}}=\hbar\omega$ induced by the 
corresponding step operator $\hat{J}_+=\hat{J}_x+i\hat{J}_y$. 
Numerically the spurious solutions decouple  
from the physical RPA solutions in a stable manner if the mean field problem is  
solved  accurately enough. In the discussion below we will only refer to the physical RPA  
solutions. They are in all cases well decoupled from the spurious solutions in our numerical procedure. 
The RPA phonon energy 
gives the energy splitting between the zero-phonon lower band and the 
excited one-phonon band at a given rotational frequency $\omega$.  
From the calculated RPA amplitudes we derive the inter-band transition 
rates using the method of~\cite{KN86}. Since the RPA does not give any contribution  
to the intra-band transition rates, we use the TAC results~\cite{Fr00} for those.

\subsection{Energies} 
\begin{figure}[htbp] 
\centerline{\includegraphics[clip,width=12.5cm]{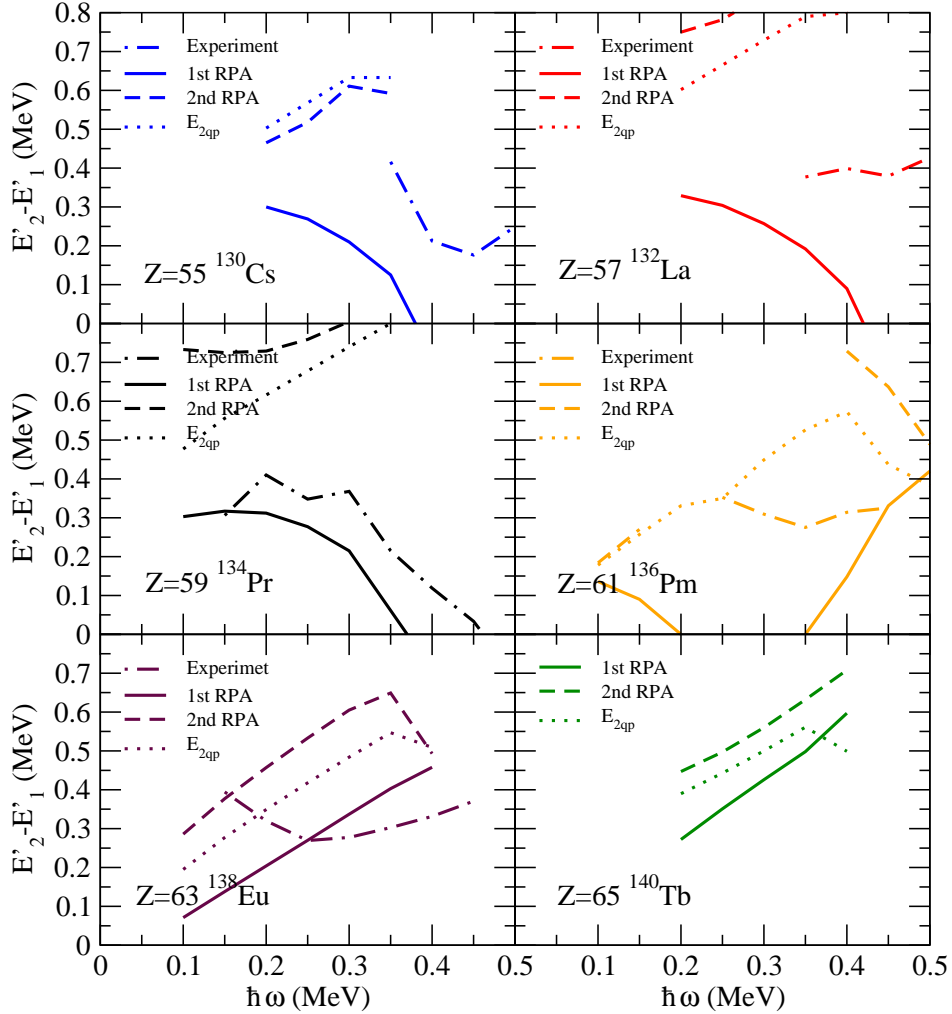}} 
  \caption{(Color online) Energy differences in the rotating frame between the two  
    chiral partner bands in the N=75 isotone chain. RPA results are compared with  
    experimental data~\cite{KS03,SC02,TA06,HR01,HB01}. We also plot the second RPA  
    phonon energy and the 2qp energy, $E_{2qp}$. The experimental  
    values show the difference between the experimental routhians of the two bands 
at grid points of $\hbar\omega$ that were calculated by interpolation.} 
  \label{fig:dEN75} 
\end{figure} 
 
\section{Results} 
\label{sec:Results}   
Below we discuss the results of the TAC + RPA calculations for the N=75 isotones and the  
Z=57 isotopes. The limits of the studied region are chosen by the following consideration.
 For lower Z or larger N we approach the shell closing where the 
deformation disappears. For smaller N the triaxiality disappears.  
For larger Z  one approaches the proton drip line, where very 
little high spin data are available.

First we solved the selfconsistent TAC problem using the Hamiltonian in  
Eq.~(\ref{eq:H1}) for the 2qp configuration $\pi h_{11/2} \otimes \nu h_{11/2}$.
The resulting s.c. values $\varepsilon, \gamma, \vartheta$ and $ \varphi$
are summarized in table \ref{tab:epsgam}.
For N=75,  the neutron chemical potential is located in the upper 
part of the $h_{11/2}$ shell, and the quasi neutron has hole character.
For Z=57, the  proton chemical potential is located at the bottom of the $h_{11/2}$ shell, and the 
quasi-proton has particle character. Along the N=75 isotone chain, the proton chemical potential approaches
the middle of the  of the $h_{11/2}$ shell at Z=65. The change of the quasi proton is reflected
by  the TAC solution. For low Z and low $\omega$, the energy minimum is at $\varphi=0$, because the particle - like
quasi proton aligns with the short axis and the hole - like quasi neutron with the long axis. The two axes
define the short - long plane $\varphi=0$.  With increasing $\omega$ the intermediate axis, which has a larger
moment of inertia, becomes progressively favored until  the $\varphi=0$ minimum
becomes unstable and static chirality sets in~\cite{Fr01,DF00,OD04,FM97}.  
The preference  of the short axis over the intermediate one  becomes attenuated
when the quasi proton loses its particle character with increasing Z. As a consequence,
the minimum moves to $\varphi=90^o$ into the intermediate-long plane.  For Z=65 the    $\varphi=90^o$ 
 minimum is stable. It is very shallow  for Z=63. In these cases the minimum becomes deeper with increasing
 $\omega$. The Z=61 case is in between. The $\varphi=0$ minimum is very shallow.
 Static chirality sets  in at the relatively low frequency of $\hbar \omega=0.20$ MeV. The chiral minimum is very weak,
 such that $\varphi=90^o$ becomes the location of the minimum for $\hbar \omega\ge0.40$MeV


In Figs.~\ref{fig:dEN75}  
and~\ref{fig:dEZ57} the energy of the lowest RPA phonon is compared with  the  
experimental energy splitting between the chiral bands. 
For reference we also plot the energy of the lowest 2qp excitation 
(relative to the  $\pi h_{11/2} \otimes \nu h_{11/2}$ TAC configuration),
as well as the energy of the second physical RPA phonon. 
 In the lighter nuclides the energy of the lowest RPA phonon is  
substantially smaller than the 2qp excitation energy and  
the energy of the second RPA 
phonon. In the heavier nuclides the difference is smaller.  
For the lowest phonon being a well developed collective excitation it 
 must have a substantially smaller energy than the next 
excitations. When the lowest phonon sits  in a region where other  
excitations are located  
the collective strength will become fragmented, and the observed bands will be much 
like 2qp excitations.  
 
Figure~\ref{fig:dEN75} shows the results for the N=75 isotones. 
The RPA results  reflect the properties of the TAC  
mean field solutions.  In the low-Z part of the A=135 region  
the TAC energy minimum sits at $\varphi=0$ at  
low spin. As we move towards larger Z the minimum moves over to $\varphi=90^\circ$.  
With increasing $\omega$, $\varphi=90^\circ$ is 
preferred, because the intermediate axis has the largest moment of inertia. In the lighter systems, 
where the TAC solutions have $\varphi=0$, 
 the phonon energy decreases with spin, reaching zero at the critical frequency  
indicating the onset of static chirality  ($\varphi>0$).  
In the heavier systems, where the TAC solutions have $\varphi=90^\circ$ already at the  
band head,  the phonon energy increases with increasing angular momentum.  
$^{136}$Pm is the limiting case where we have a chiral vibration  
around $\varphi=0$ and decreasing phonon energy at low spin, that goes to 
a static chiral regime  with $0<\varphi<90^\circ$, and finally a second chiral 
vibration around $\varphi=90^\circ$ and increasing phonon energy. 
Comparing with experiment,   $^{130}$Cs,  
$^{134}$Pr, $^{138}$Eu have the right trend and magnitude of the energy  
splitting. $^{132}$La, $^{136}$Pm have a much more constant 
energy splitting in the experiment than obtained by the RPA calculations.

\begin{figure}[htbp] 
\centerline{\includegraphics[clip,width=6.25cm]{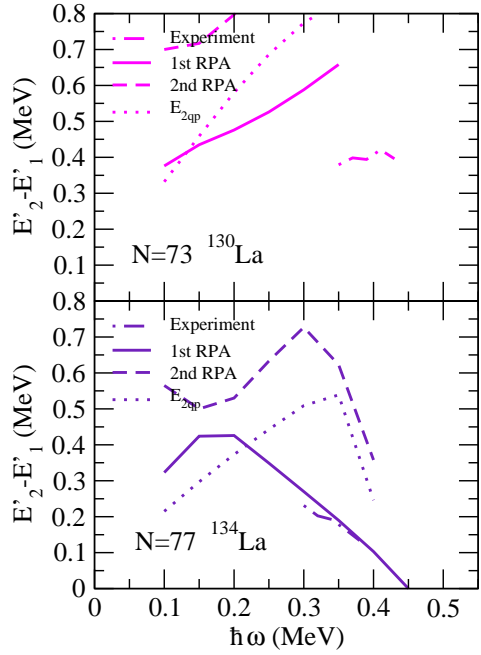}} 
  \caption{(Color online) Energy differences in the rotating frame between the two  
    chiral partner bands in Z=57 isotopes. RPA results are compared with experimental  
    data~\cite{KS01,BB01}. We also plot the second RPA phonon energy  
    and the 2qp energy, $E_{2qp}$.} 
  \label{fig:dEZ57} 
\end{figure} 

\begin{figure}[htbp] 
\centerline{\includegraphics[clip,width=12.5cm]{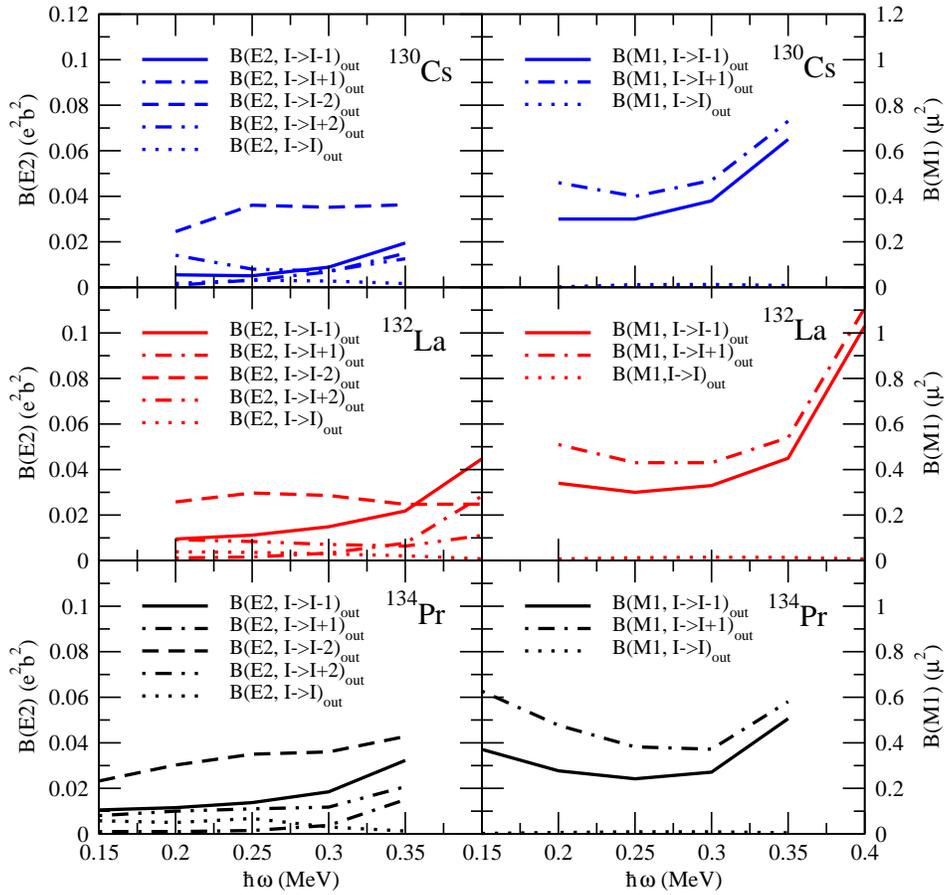}} 
  \caption{The RPA inter-band transition rates in $^{130}$Cs, $^{132}$La and  
    $^{134}$Pr.} 
  \label{fig:TRrpaN75l} 
\end{figure} 
\begin{figure}[htbp] 
\centerline{\includegraphics[clip,width=12.5cm]{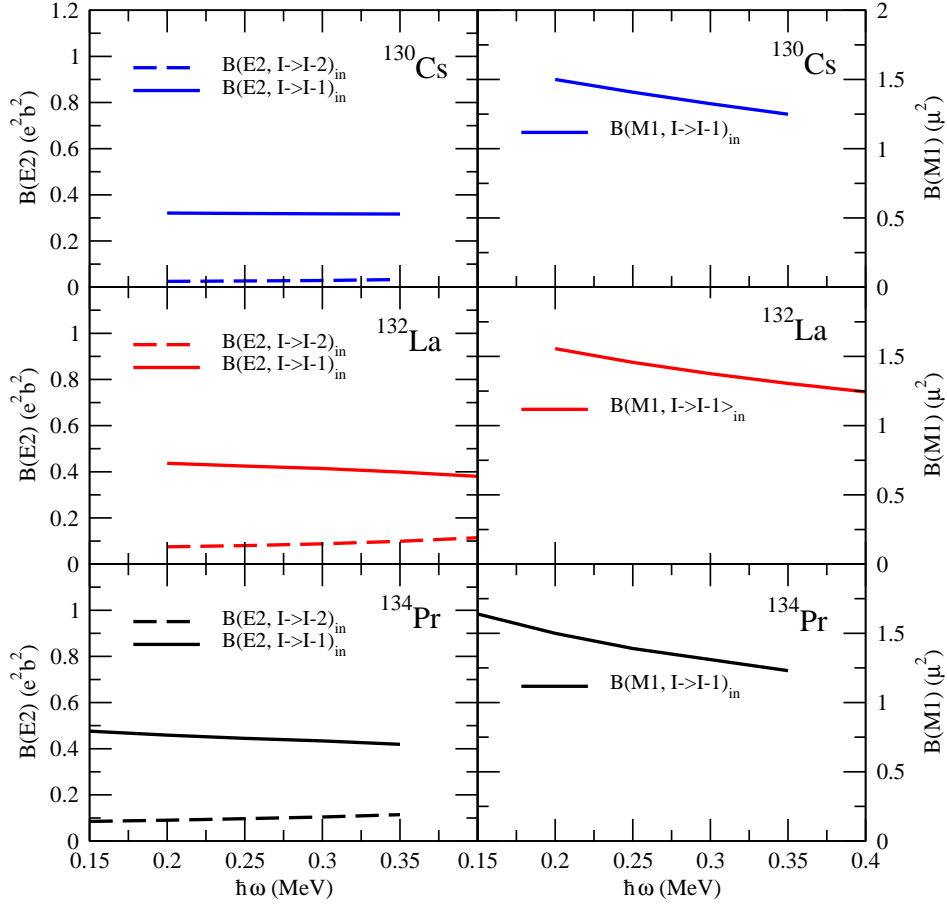}} 
  \caption{The TAC intra-band transition rates in $^{130}$Cs, $^{132}$La and  
    $^{134}$Pr.} 
  \label{fig:TRtacN75l} 
\end{figure} 
The inter-band M1 transitions are much stronger in the lighter N=75 isotopes 
where the B(M1) values are typically around 1/3 of the intra-band values. In the heavier 
N=75 isotopes as well as in $^{134}$La the inter-band B(M1) are weak. 
\begin{figure}[htbp] 
\centerline{\includegraphics[clip,width=12.5cm]{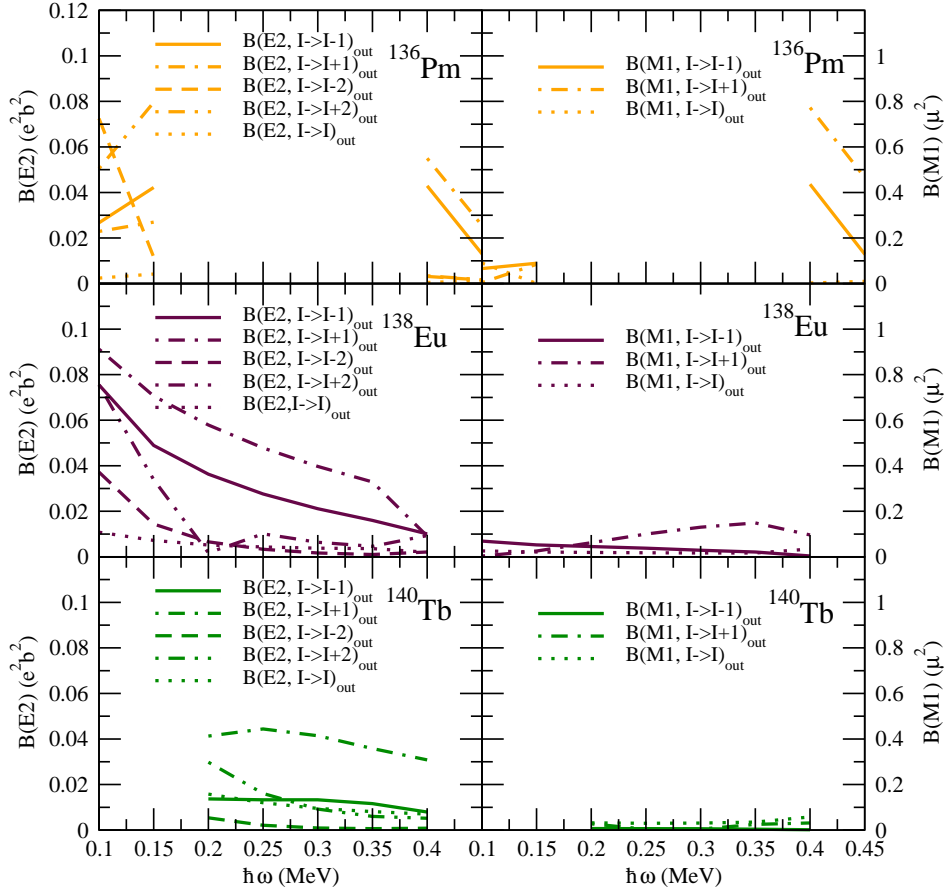}} 
  \caption{The RPA inter-band transition rates in $^{136}$Pm, $^{138}$Eu and  
    $^{140}$Tb.} 
  \label{fig:TRrpaN75u} 
\end{figure} 
\begin{figure}[htbp] 
\centerline{\includegraphics[clip,width=12.5cm]{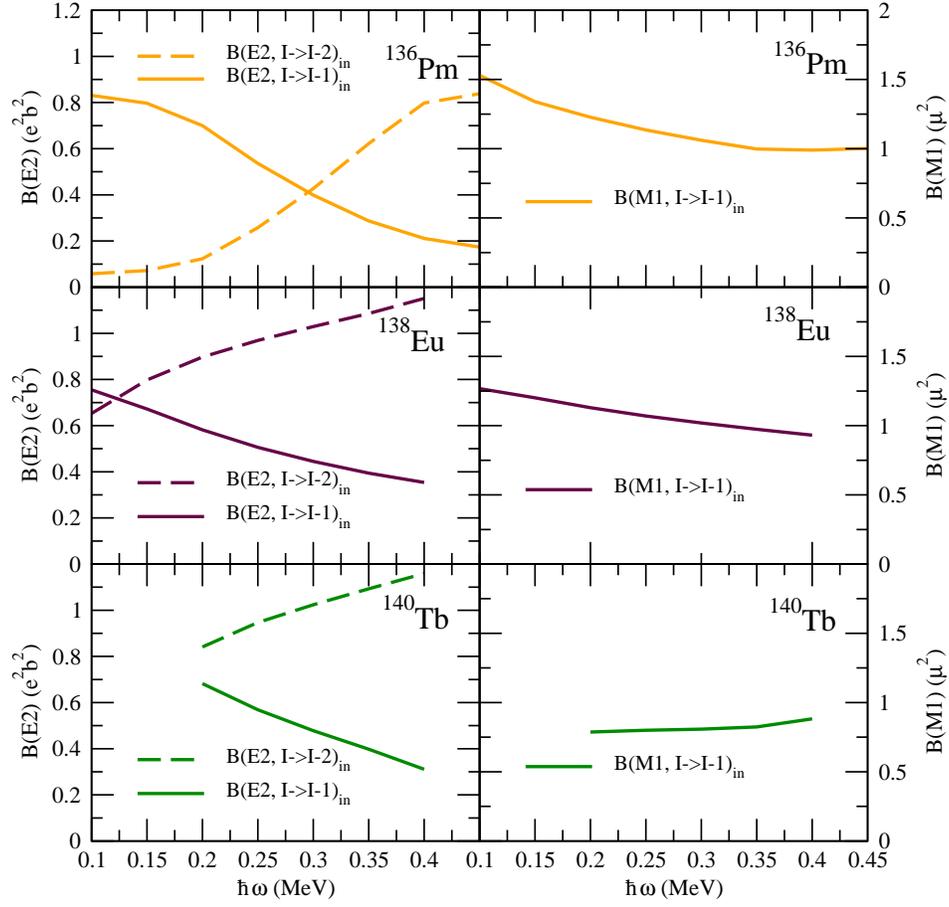}} 
  \caption{The TAC intra-band transition rates in $^{136}$Pm, $^{138}$Eu and  
    $^{140}$Tb.} 
  \label{fig:TRtacN75u} 
\end{figure} 

Figure~\ref{fig:dEZ57} show the results for the Z=57 isotopes. The magnitude  
of the energy splitting is reproduced but there is a problem with the   
frequency trend in $^{132}$La. In experiment $^{132}$La looks similar to $^{130}$La 
while in the calculations it looks more like $^{134}$La. 
 
Our TAC+RPA calculations  give the correct energy splitting away from the  
region where the RPA energy approaches zero, which is  where we can  
expect the RPA to work well. However, spin dependence is not always well  
reproduced. The RPA phonon energy tends to approach zero faster and more often than  
in experiment, which indicates that anharmonic effects are needed to  
understand the data. 
The RPA phonon energy is generally increasing or decreasing with rotational frequency. 
The experimental energy splitting, on the other hand, does not change much 
with $\omega$ in some cases, which seem to appear predominantly in the  
frequency regions where the TAC gives static chirality,  
$0<\varphi<90^\circ$, i.e. TAC + RPA predicts zero splitting. The 
constant energy splitting in these regions should be attributed to the  
tunneling between the left- and right-handed solutions. Its relatively  
large value of about $300$ keV indicates strong mixing between  
the two chiral configurations,  
which is consistent with the flat potential in $\varphi$ direction  
in Fig.~\ref{fig:pr134PEStf} and our analysis of the composition of the RPA phonon 
wave function discussed in section~\ref{sec:amplitude} below. In $^{134}$Pr the two bands  
cross over in experiment. In $^{130}$Cs and $^{135}$Nd (not shown here, see~\cite{MA07}) 
one sees in experiment an avoided crossing with some interaction. In all three cases, 
the crossing frequency correlates with the frequency where the phonon energy goes to zero.    
At moment, it remains unclear why these nuclei are different from 
$^{132}$La and $^{136}$Pm with a nearly frequency-independent 
energy difference between the bands. 
 To understand this phenomenon further it is clearly necessary to take into account 
large amplitude anharmonic effects, which will be done in a forthcoming 
publication~\cite{ADF11}.

\subsection{Transition rates} 
In addition to the  energy splitting, the RPA also provides the transition rates between  
the two chiral partner bands which are connected by E2 and M1 radiation with spin differences
$\Delta I=0,\pm 1, \pm 2$ . In accordance with the RPA the transition amplitudes are calculated 
with the multipole operators $\hat{\cal M}_k(\lambda;P)$  in the P-system 
where $\lambda$ defines the multipole order and $k$ the 3-component of the transition. In detail 
the transition operators are given by
\begin{eqnarray}\label{eq:BE2} 
  \hat{\cal M}_k(E2;L) &=& e_p r_p^2 {Y}_{2k} (p)+e_n r_n^2 {Y}_{k} (n),
 \qquad\qquad\qquad k=0, \pm 1, \pm 2 \\
  \hat{\cal M}_k(M1;L)& =&\frac{3}{4\pi}
g^{(l)}_p\hat{l}_{1k}(p)+g^{(s)}_p \hat{s}_{1k}(p)+g^{(s)}_n\hat{s}_{1k}(n) 
,\quad\quad k=0, \pm 1
\label{eq:BM1}
\end{eqnarray} 
The effective charges for the E2 are  $e_p= 1+Z/A$ for protons  and $e_n=Z/A$ for 
neutrons. The orbital g-factor for M1 is $g^{(l)}_p= 1 \mu_N$ for protons  and 0 for 
neutrons. The spin g-factors $g^{(s)}_p$ and $g^{(s)}_n$ are 0.7 times the values for the 
free proton or neutron. The above multipole operators are treated by boson expansion 
in linear order only. Finally we have to remember  that the radition is observed in the 
laboratory system which is taken into account by rotating $\hat{\cal M}_k(\lambda;P)$ 
from the P- to the L-frame: 
\begin{equation} 
  \hat{\cal M}_m(\lambda;L) = \sum_k D^\lambda_{k m }(\varphi,-\vartheta,0)\hat{\cal M}_k(\lambda;P).
\end{equation} 
The expression applies to the different planar TAC solution corresponding to $\varphi=0$, 
i.e that angular momentum vector lies in the plane spanned by the long and short axes, 
and to $\varphi=90^o$,  i.e that angular momentum vector lies in the plane spanned 
by the long and intermediate axes, which gives the additonal phase factor $e^{-ik\frac{\pi}{2}}$ 
in the sum.   

The component $m$ determines the actual spin difference of the transition $I\to  I-m$.  Hence we
obtain the reduced transition probabilities for the inter-band transitions 
 \begin{equation}\label{eq:BE2b} 
  B(E2,I\rightarrow I-m) = \left|\left< 1\left|\hat{\cal M}_m(E2;L) \right|0 \right>\right|^2
  \quad\text{and }\quad
 B(M1,I\rightarrow I-m) = \left| \left< 1\left|\hat{\cal M}_m(M1;L) 
	\right| 0 \right> \right|^2 
\end{equation} 
where $\left| 0 \right>$ is the RPA ground state and $\left| 1 \right>$ is the 
first excited RPA state at the rotational frequency $\omega=\omega(I)$.  

	In Fig.~\ref{fig:TRrpaN75l}-\ref{fig:TRtacZ57} we  
show the resulting B(E2) and B(M1) inter-band rates as a function of the frequency $\omega$.
 The intra-band values calculated  
by means of  TAC~\cite{Fr00} are also included. 

For the E2 transitions in the lighter N=75 isotopes the dominant inter-band  
transition is $I\rightarrow I-2$, except when approaching  
the critical frequency  the transition $I\rightarrow I\pm1$ increases in strength.  
In the heavier N=75 isotopes  $B(E2,I\rightarrow I+1)$ becomes the largest 
inter-band value. However, the $I\rightarrow I+1$ or $I+2$  
will often not be seen in experiment  due to their low or negative transition energy.  
\begin{figure}[htbp] 
\centerline{\includegraphics[clip,width=12.5cm]{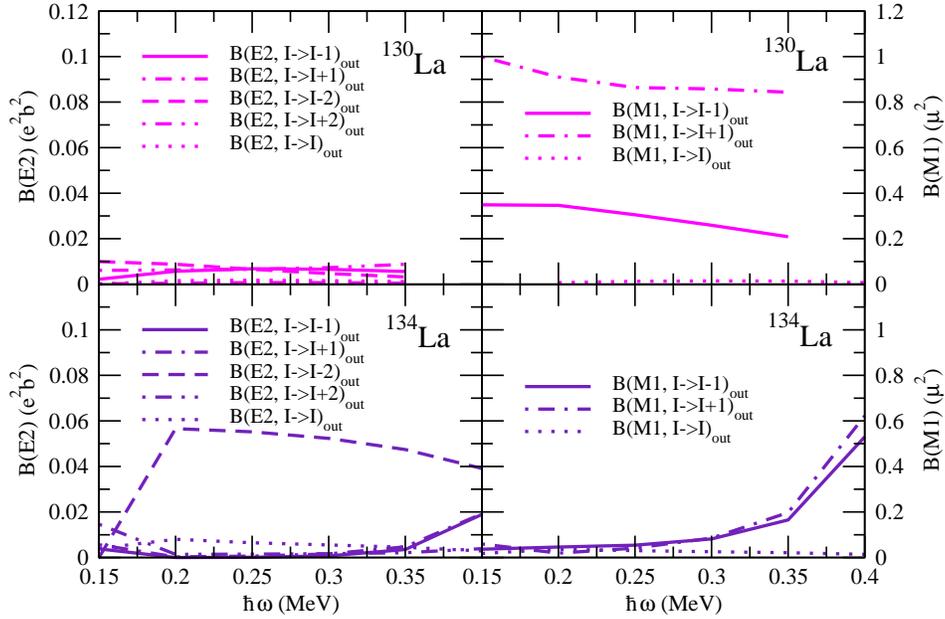}} 
  \caption{The RPA inter-band transition rates in $^{130}$La and $^{134}$La.} 
  \label{fig:TRrpaZ57} 
\end{figure} 
\begin{figure}[htbp] 
\centerline{\includegraphics[clip,width=12.5cm]{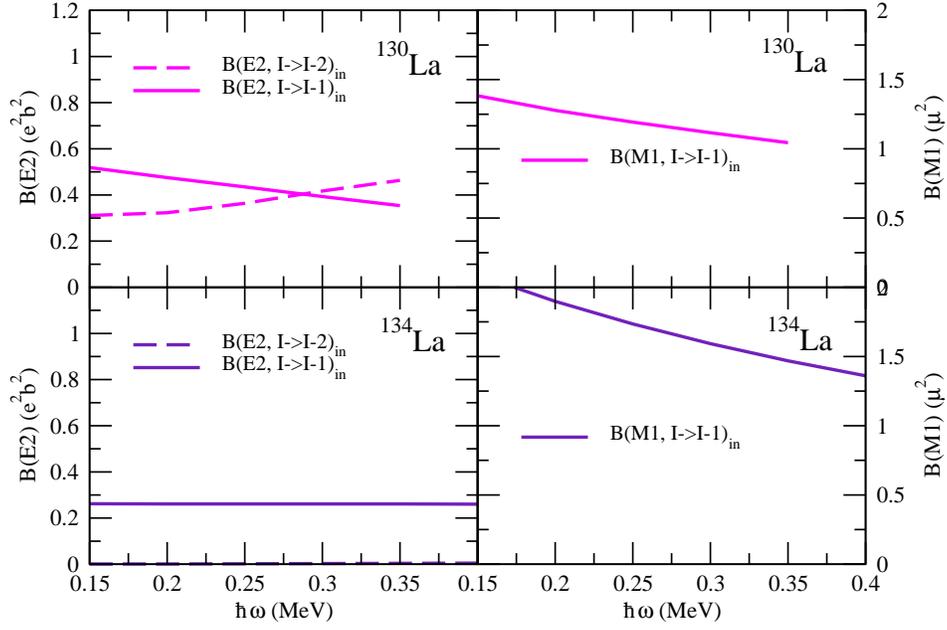}} 
  \caption{The TAC intra-band transition rates in $^{130}$La and $^{134}$La.} 
  \label{fig:TRtacZ57} 
\end{figure} 
The ratios of the different transition probabilities vary substantially over the  
studied region. The $B(E2,I\rightarrow I\pm 1)$ and $B(M1,I\rightarrow I\pm 1)$  
are typically the strongest. 
 
There are experimental data for the inter-band transition rates in  
$^{134}$Pr~\cite{TA07}. Our  calculated rates are somewhat larger than the ones 
seen in experiment. 
There are also data available for the ratio of $B(M1)_{out}/B(M1)_{in}$ in  
$^{136}$Pm~\cite{HR01}, but unfortunately not in the frequency range where we have RPA  
results.

\section{Shape and orientation oscillation amplitudes} 
\label{sec:amplitude} 
The energies and the transition rates do not give direct  
information about the character of the RPA phonon. 
The question is, if it represents predominantly oscillations of the 
orientation of the quadrupole tensor relative to the angular momentum vector,
which we then classify as chiral vibrations, or predominantly oscillations of 
the deformation parameters, or a combination of both types of oscillation.  
From the curvature of potential energy surfaces 
as seen in Fig.~\ref{fig:pr134PESgf} and~\ref{fig:pr134PEStf}
one can only make some qualitative guesses.
In the following  we determine quantitatively the size of 
shape and angle oscillations of the RPA phonon.

The RPA describes quantized oscillations of the quadrupole tensor relative to the 
uniformly rotating quadrupole tensor of the self-consistent TAC solution. 
As well known, the explicit
time dependent state $|t\rangle$ corresponding to TAC+RPA is the small amplitude periodic solution of the time
dependent mean field problem (TDHF, cf. e.g. \cite{Negele82}). 
The TAC solutions are found in the P-coordinate system, which uniformly rotates about the z-axis of the
L-system.  Within the P-frame, the quadrupole
tensor of the TAC solution does not depend on time, and  $\langle \hat Q_{\pm 1}\rangle$=0 and  
$\langle \hat Q_{2}\rangle$=$\langle \hat Q_{-2}\rangle$. The time dependence is generated  by the
RPA correlations, which are reflected by transition matrix elements   of the quadrupole tensor.
It is useful to consider the set $\hat Q_{_{k\pm}}$ of hermitean quadrupole 
operators defined by 
\begin{eqnarray}
\label{combin}
  \hat Q_{_{1+}} &=& \frac{\hat Q_{_{1}}+ \hat Q_{_{-1}}}{i\sqrt{2}},
\quad\hat Q_{_{1-}} = \frac{\hat Q_{_{1}}- \hat Q_{_{-1}}}{\sqrt{2}},\nonumber\\
   \hat Q_{_{2+}} &=& \frac{ \hat Q_{_2}+ \hat Q_{_{-2}}}{\sqrt{2}},\quad
  \hat Q_{_{2-}} = \frac{ \hat Q_{_2}- \hat Q_{_{-2}}}{i\sqrt{2}}.
\end{eqnarray}
As discussed in the Appendix, their mean values oscillate with the frequency  $ \Omega$ 
corresponding to  RPA phonon excitation energy $E_{_{RPA}}=\hbar\Omega$, and the TDHF state can be normalized such  
that the amplitudes are equal to the  transition matrix elements, i. e.
\begin{equation}\label{qfluc} 
\langle t| \hat Q_{_{k\pm}}|t\rangle - \langle \hat Q_{_{k\pm}}\rangle 
= \langle 1 |{\hat Q}_{_{k\pm}} |0\rangle \cos \Omega t \equiv\Delta Q_{_{k\pm}} \cos \Omega t .
\end{equation}

The quadrupole oscillations can be decomposed into oscillations of 
the shape parameters $\varepsilon$ and $\gamma$ ($\beta$ and $\gamma$ vibrations, respectively) 
and into oscillations of the orientation 
angles $\vartheta$, $\varphi$, and $\psi$  of the principal axes of quadrupole tensor with respect to
the uniformly rotating P-system. That is, we derive a new set of transition
amplitudes  $\Delta \vartheta$, $\Delta \varphi$, and $\Delta \psi$, representing the 
chiral vibrational part of the excitation, and $\Delta \varepsilon/\varepsilon$ and $\Delta\gamma$
representing the shape oscillation part.  
Exploiting the shape selfconsistency conditions (\ref{sc}), 
\mbox{$\langle \hat Q_{_0}\rangle = C\, \varepsilon\cos{\gamma}$} and 
\mbox{$\langle \hat Q_{_{2+}}\rangle= -C \,\varepsilon\sin{\gamma}$},  
where \mbox{$C=2/3\,\hbar\omega_{_0}/\kappa_{_0}$}, we derive  in the  Appendix
\begin{eqnarray}
\frac{\Delta\varepsilon}{\varepsilon}
&=&\frac{\langle \hat Q_{_0}\rangle\Delta Q_{_0} +
\langle \hat Q_{_{2+}}\rangle\Delta Q_{_{2+}}}
{\langle \hat Q_{_0}\rangle^2+\langle \hat Q_{2+}\rangle^2},\label{shapefluc1}\\ 
\Delta\gamma
&=& \,\frac{\langle \hat Q_{_{2+}}\rangle\Delta Q_{_{0}}-
\langle \hat Q_{_{0}}\rangle\Delta Q_{_{2+}}}
{\langle \hat Q_{_0}\rangle^2+\langle \hat Q_{2+}\rangle^2}\label{shapefluc2}. 
\end{eqnarray}
Considering the following small rotations 
\begin{eqnarray}
R_3&=& 1-i\Delta\varphi \hat J_3,\nonumber\\
R_2&=&1-i\Delta\vartheta \hat J_2,\nonumber\\
R_z&=& 1-i\Delta\psi \hat J_z=1-i\Delta\psi(\cos{\vartheta}\hat J_3+\sin{\vartheta}\hat J_1),
\end{eqnarray}
where the angular momentum operators are taken in RPA order,
and evaluating the expectation values $\langle R_i\hat Q_{_{k}}R_i\rangle$ 
  we derive in the Appendix 
\begin{eqnarray}
\Delta\varphi
&=&\frac{\Delta Q_{_{2-}}}{2\langle  \hat 
Q_{_{2+}}\rangle}-\Delta\psi\cos{\vartheta}, \label{anglefluc1} \\
\Delta\vartheta
&=&\frac{\Delta Q_{_{1-}}}{\langle \hat Q_{_{2+}}\rangle-
\sqrt{3}\langle \hat Q_{_0}\rangle},\label{anglefluc2} \\
\Delta\psi
&=&\frac{\Delta Q_{_{1+}}}{\,\sin{\,\vartheta }\,
(\langle \hat Q_{_{2+}}\rangle+\sqrt{3}\langle \hat 
Q_{_0}\rangle)}.\label{anglefluc3}
\end{eqnarray}
Similar equations  were derived by 
Shimizu and Matsuzaki \cite{SM96}  (cf. their Eqs.(4.21)) for the  wobbling mode, which correspond to
the special case  $\vartheta =0$ where $\Delta\psi$ becomes zero (see Appendix). 
\begin{figure}
\centerline{\includegraphics[clip,width=12.5cm]{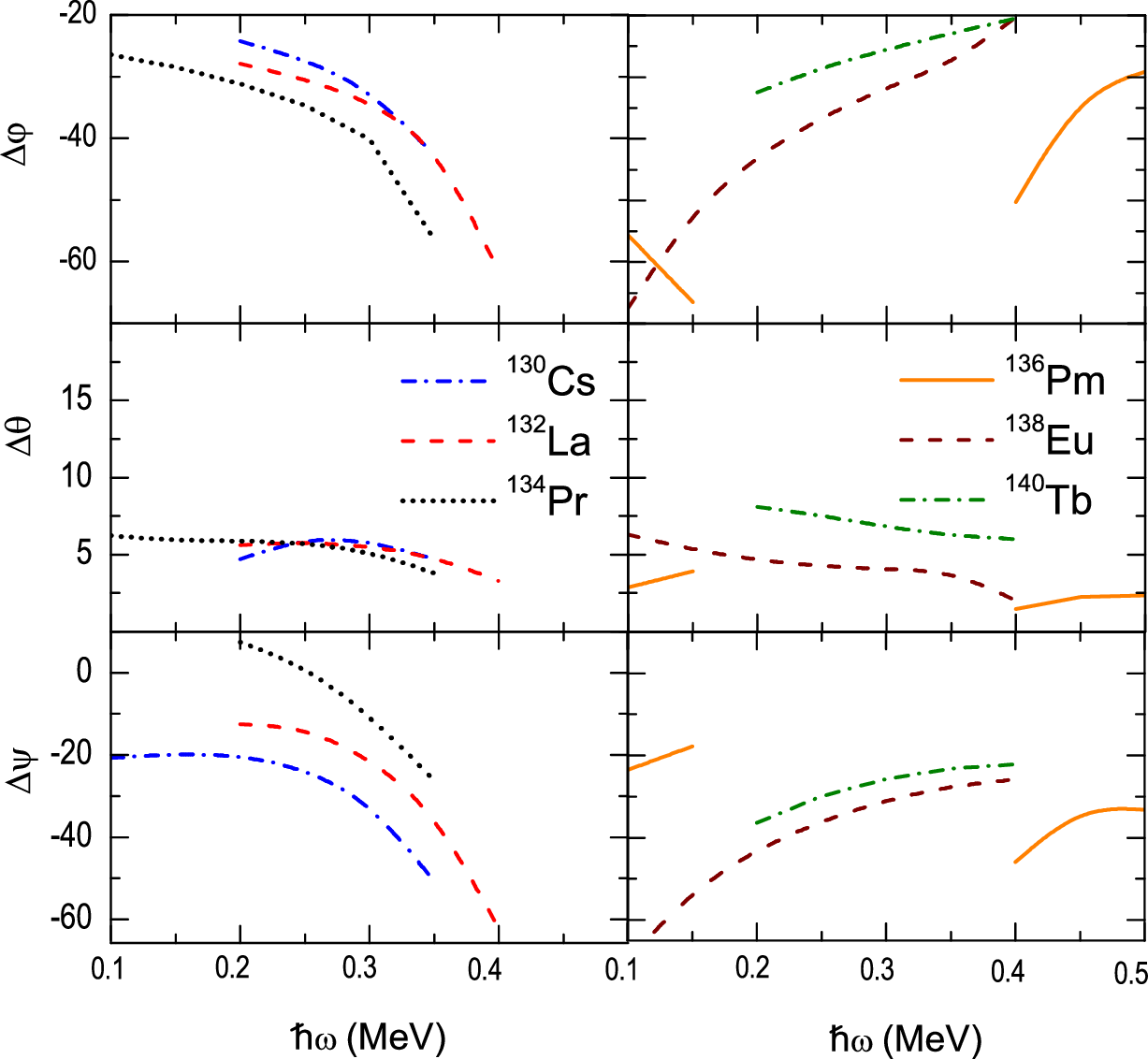}} 
  \caption{(Color online) The amplitudes of the Euler angles in degrees in the lowest RPA phonons
of the N=75 isotone chain. Cs, La and Pr in the left panels and Pm, Eu and Tb  
    in the right panels.} 
  \label{fig:AngleN75} 
\end{figure}
\begin{figure}[htbp] 
\centerline{\includegraphics[clip,width=12.5cm]{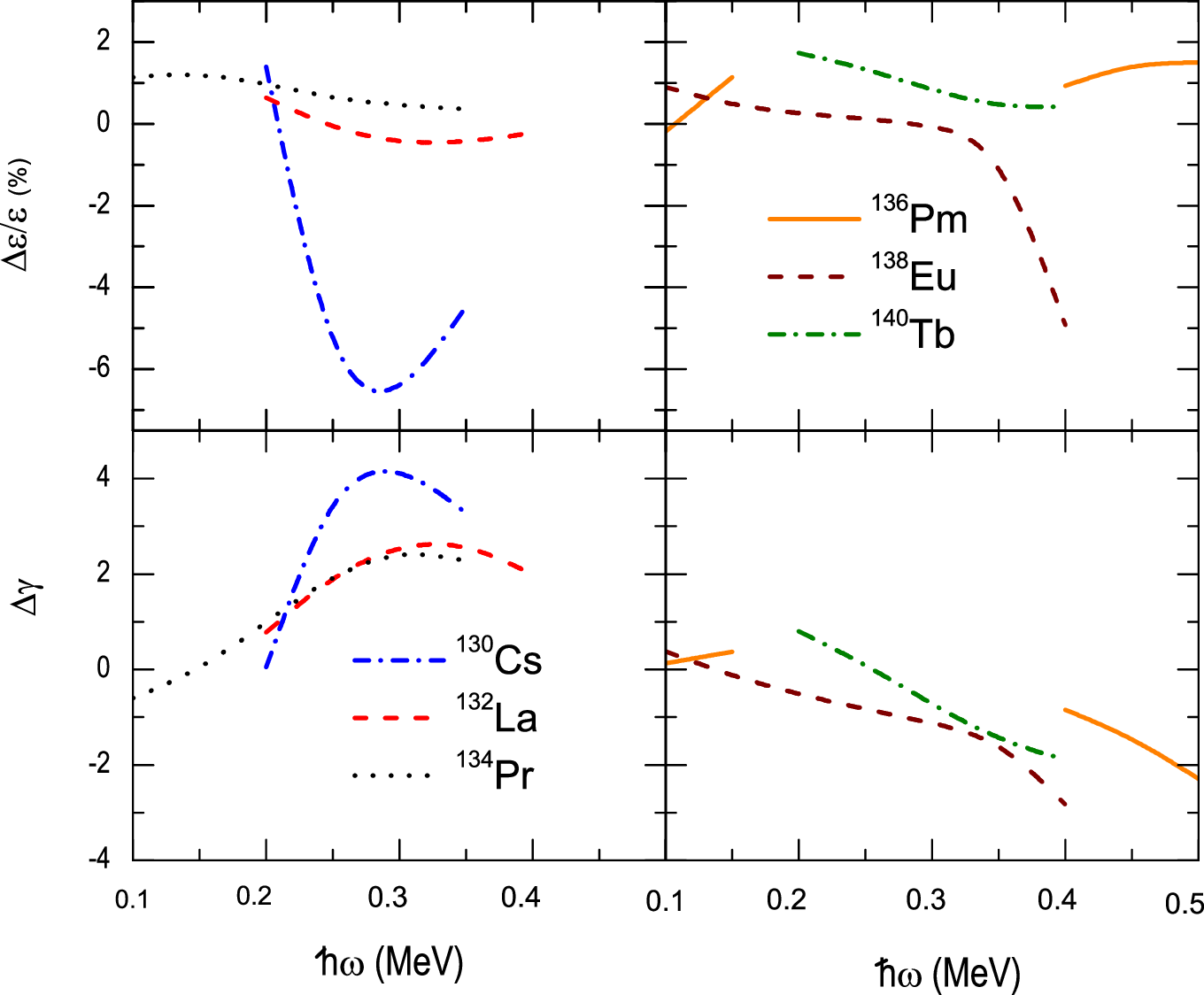}} 
  \caption{(Color online) The amplitudes of shape parameters in the lowest RPA phonons  
    in the N=75 isotone chain ($\gamma$ in degrees). Cs, La and Pr in the left panels and Pm, Eu and Tb  
    in the right panels.} 
  \label{fig:ShapeN75} 
\end{figure} 

\begin{figure}[htpp]
\begin{minipage}[b]{0.47\linewidth}
\centering
\centerline{\includegraphics[clip,width=6.25cm]{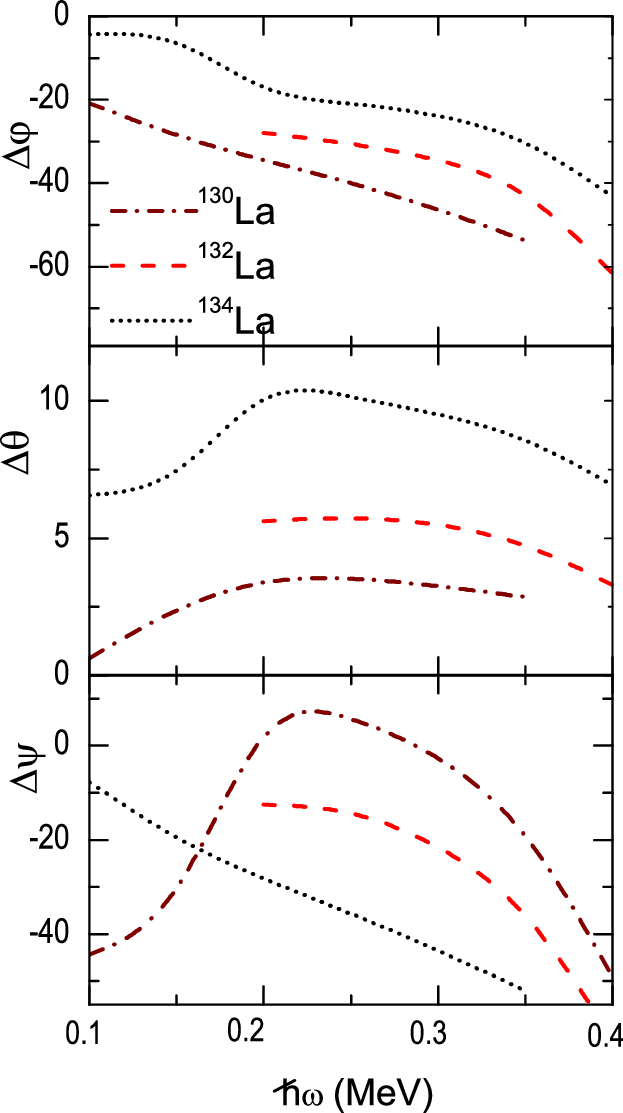}} 
  \caption{(Color online) The amplitudes of Euler angles in degrees in the lowest RPA phonons in  
    the Z=57 isotones.} 
  \label{fig:AngleZ57} 
\end{minipage}
\hspace{0.5cm}
\begin{minipage}[b]{0.47\linewidth}
\centering
\centerline{\includegraphics[clip,width=6.25cm]{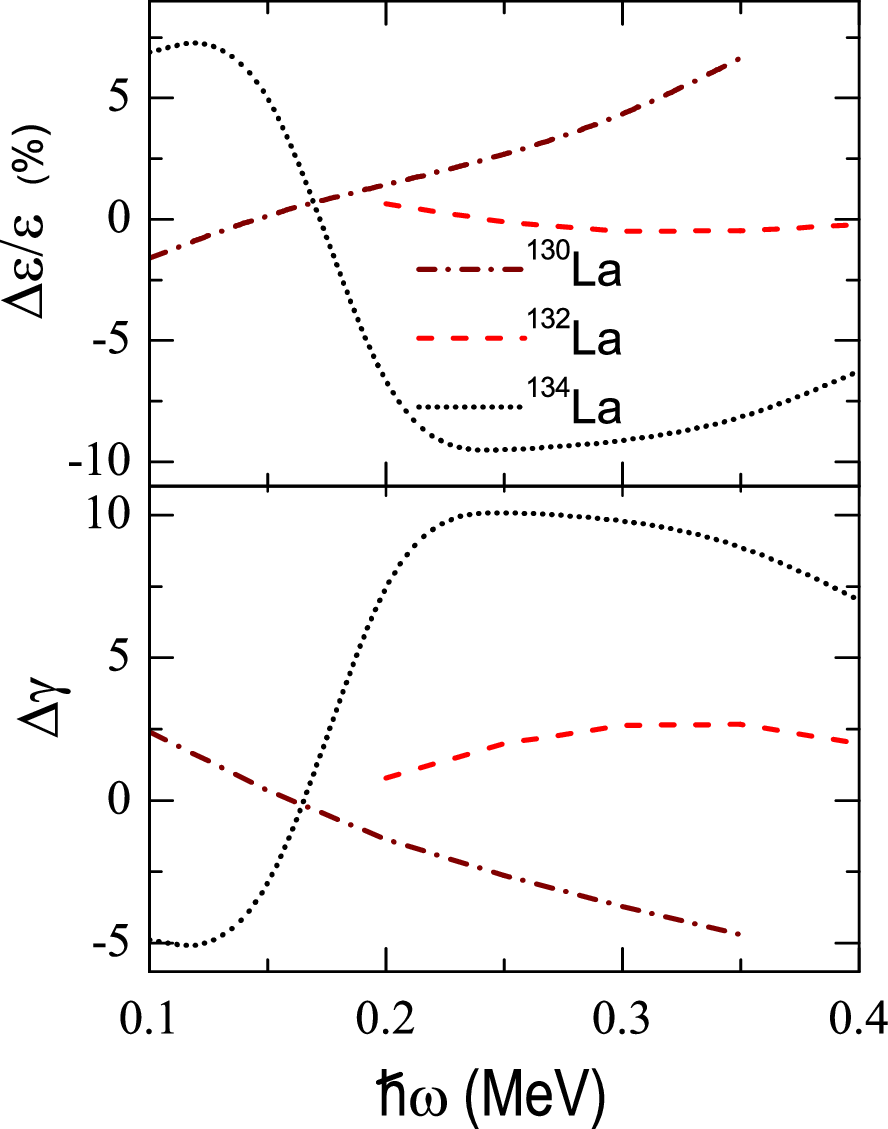}} 
  \caption{(Color online) The amplitudes of shape parameters in the lowest RPA phonons  
    in the Z=57 isotones ($\gamma$ in degrees).} 
  \label{fig:ShapeZ57} 
\vspace*{2.8cm}
 \end{minipage}
\end{figure}

The results are shown in Fig.~\ref{fig:AngleN75} and~\ref{fig:ShapeN75}. One  
can clearly see that the lowest phonon  is dominated by 
orientation oscillations, especially in the $\varphi$ direction, which is consistent  
with the TAC potential energy surface in Fig.~\ref{fig:pr134PEStf}. The amplitude  
$\Delta \varphi$ starts at  30-40$^\circ$ and increases toward    
 the point where 
the RPA energy goes to zero. It diverges at the transition to aplanar chiral rotation. 
As discussed above, this can happen with increasing or decreasing $\omega$, depending on the nuclide.
The amplitudes     $\Delta\psi$ behave in the same way,  being somewhat smaller than $\Delta \varphi$.
The amplitude of the  $\vartheta$-oscillations is  5-10$^\circ$. 
In all cases, the shape oscillations are small in amplitude compared to the orientation  
oscillations. The amplitudes $\Delta\varepsilon$ are 
 few percent of the equilibrium values and   
$\Delta \gamma <6^o$ for all cases but $^{134}$La (cf. next paragraph).  
 This is in contrast 
to Ref.~\cite{TA06}, which,  taking into account the shape degrees 
of freedom in the framework of the the IBAFF two-particle-core approach, 
found a strong coupling between the shape and orientation  
degrees of freedom. In our microscopic approach, the lowest phonon is 
a rather clean chiral vibration.  
  
 Figures~\ref{fig:AngleZ57} and~\ref{fig:ShapeZ57} show  
how the structure of the lowest RPA phonon develops when   
moving away from the  N=75 chain.   In the N=73 nuclide $^{130}$La, the angle oscillations
also dominate although   
$\gamma$ vibration becomes more important at large $\omega$.  This is consistent with a
reduced  stability of the triaxial deformation, which eventually disappears  with decreasing N.
 In $^{134}$La with N=77  we are approaching the N=82 shell closure, which leads to a  
reduction in the deformation and increased amplitudes of the shape oscillations 
in the RPA phonon. The case of $^{134}$La can no longer be thought of as a chiral 
vibration even though it has large components of orientation oscillations in the  
wave function. The large jump in several of the $^{134}$La curves is due to the 
level crossing between the first and second RPA solution seen in Fig.~\ref{fig:dEZ57}.

\section{Conclusions} 
\label{sec:Conclusions} 
We have combined the TAC mean field approach with the RPA for a modified Quadrupole-Quadrupole interaction
in order to calculate  the energy splitting between chiral partner  
bands, which is given by the excitation energy of the lowest RPA solution.
Our TAC+RPA calculations also give the intra- and inter-band transition rates.  

By analyzing the RPA amplitudes we found that near Z=57 and N=75  
the lowest phonon is a rather pure  collective  
chiral vibration. It represents  a slow oscillation of the quadrupole shape relative to the angular momentum vector, which
takes wide excursions into the left - and right - handed arrangements of the three principal axes and the angular momentum. The amplitudes of the shape oscillations are found to be much smaller  
than the ones of the orientation oscillations.   
The large amplitude nature of the angle oscillations indicate that 
for a full understanding of the chiral bands one has to take into account  
effects that go beyond the harmonic RPA approximation in the orientation  
degrees of freedom. The anharmonicities will change the collective mode in a qualitative way  where the 
energy of the first RPA solution goes to zero and the mean field attains static chirality. 
This transition is encountered for most of the studied cases. Since our investigation
indicates that the shape degrees are well decoupled from the angle ones,
it seems promising to try to describe the transition in terms of an effective 
Hamiltonian depending only on the components of the total angular momentum~\cite{ADF11}.

Our TAC+RPA calculations give phonon energies in the order of 300 keV or lower, which is comparable with the 
experimental energy differences between the chiral partner bands. However the  trend with 
increasing angular momentum is not always  reproduced. The calculated phonon energy 
goes to zero where the TAC solution attains static chirality.  One would expect 
TAC+RPA to describe the angular momentum dependence of the energy difference between  
the partners in a qualitative manner, i.e. reproduce if it is increasing a or decreasing and   
give zero splitting for static chirality of the TAC solution instead 
of a finite small splitting seen in experiment. This is the case for $^{134}$Pr, $^{130}$Cs, $^{134}$La,
and $^{135}$Nd \cite{MA07}, whose phonon energy decreases with angular momentum, and
$^{138}$Eu, whose phonon energy increases. However, for $^{130,132}$La and $^{136}$Pm the experimental
energy difference between the partner bands is nearly constant  300-400 keV in contrast to the calculations.
The reason for the discrepancy is unclear, and it remains to be seen whether 
a large amplitude description of the chiral mode will be able to account for the experiment.
It is also noted that for a number of studied nuclides only few members of the chiral partner band are observed,
such that the angular momentum dependence of the phonon energy is not well established. 
This information would be crucial for a deeper understanding of the onset of chirality.

The calculation of the inter-band transition rates shows that the ratios of the 
different $\Delta I$ components of the transitions vary relatively fast with N and Z. 
We could not discern a systematic pattern indicating the onset of chirality.

The quite pure chiral nature of the RPA solution seems to be localized around the N=75  
isotones. Moving to more neutron-rich nuclei structure of the first RPA  
phonon changes to a complex mixture of shape and orientation degrees of freedom, 
reflecting the approach of the shell closure at N=82. Towards the more neutron deficient  
nuclides the deformation is increasing but the triaxiality is  
decreasing, which reduces the collectivity of the lowest  
RPA solutions that takes on 2qp character.

\section*{Acknowledgment} 
This work was supported by DOE grant DE-FG02-95ER4093 and the ANL-ND Nuclear Theory  
Institute.  

\appendix
\section{ }
We consider small-amplitude vibrations about a planar TAC solution in the P-system.
The solution of the TDHF equations results in oscillating quadrupole moments 
\begin{equation}
\hat Q_{_{k}}(t)=\langle\hat Q_{_{k}}\rangle+A\Delta Q_{_{k}}\cos{\Omega t}, ~~k=0,1\pm,~2\pm,
\end{equation}  
where we refer to the hermitean combination $\hat Q_{_{k\pm}}$ defined by Eqs.~(\ref{combin}). 
The periodic TDHF solutions are related  to RPA solutions 
(\ref{qfluc}) (e. g. see Ref.  \cite{Negele82} Eqs. (3.37), (3.38)) such that one has
\begin{equation}
\label{ampl}
E_{_{RPA}}=\hbar\Omega,\qquad\Delta Q_{_{k}}= \langle 1 | \hat{Q}_{_{k}} | 0 \rangle
\end{equation} 
where $|1\rangle$ corresponds to the lowest RPA excitation.  The amplitude of the oscillations
is not determined by the TDHF equations. We chose it such that $A=1$.

The oscillations of the quadrupole tensor $ \hat Q_{_{k}}(t)$ can be expressed as oscillations
of the shape parameters $\varepsilon$ and $\gamma$ and oscillations of the Euler angles 
$\varphi$, $\vartheta$ and $\psi$ which determine the orientation of the principal axes of the
oscillating quadrupole tensor relative to the P-frame.  

We start the derivations with the  shape amplitudes $\Delta\varepsilon$ and $\Delta\gamma$.
The self consistency conditions (\ref{sc})  state 
\begin{eqnarray}
\langle \hat Q_{_0}\rangle \,&=&\,\, C\, \varepsilon\cos{\gamma}\nonumber,\\
\langle \hat Q_{_{2+}}\rangle&=& -C \,\varepsilon\sin{\gamma}.
\end{eqnarray}
Variation of  both
 sides gives
\begin{eqnarray}
\label{var1}
\Delta Q_{_0}\,\, &=&\,\,\, C\, (\Delta\varepsilon\cos{\gamma}
-\varepsilon\sin{\gamma}\Delta\gamma )\nonumber,\\
\Delta Q_{_{2+}}&=& -C \,(\Delta\varepsilon\sin{\gamma}+
\varepsilon\cos{\gamma}\Delta\gamma ).
\end{eqnarray}
Inverting yields the desired relations for the shape amplitudes 
\begin{eqnarray}
\frac{\Delta\varepsilon}{\varepsilon}
&=&\frac{\langle \hat Q_{_0}\rangle\Delta Q_{_0} +
\langle \hat Q_{_{2+}}\rangle\Delta Q_{_{2+}}}
{\langle \hat Q_{_0}\rangle^2+\langle \hat Q_{2+}\rangle^2},\label{shapefluceps}\\ 
\Delta\gamma
&=& \,\frac{\langle \hat Q_{_{2+}}\rangle\Delta Q_{_{0}}-
\langle \hat Q_{_{0}}\rangle\Delta Q_{_{2+}}}
{\langle \hat Q_{_0}\rangle^2+\langle \hat Q_{2+}\rangle^2}\label{shapeflucgam}. 
\end{eqnarray}

Now we perform rotations $R_{2,3}$   with small angles  $\Delta\vartheta$ and 
$\Delta\varphi$ about the respective
axes 2 and 3 of the P-frame and a 
rotation $R_z$ with a small angle $\Delta\psi$ about the  $z$-axis of the L-frame 
(direction of angular momentum).
 We calculate the change of the quadrupole moments generated by the  reoriention, i. e.
 we take the expectation values with respect to 
the states $R_i|TAC\rangle$.
Inverting these relations, will provide us with the angles $\Delta\varphi, \Delta\vartheta, 
\Delta\psi$ of  the principal axes of $\hat Q_k(t)$ within the P-system.    
Expanding the corresponding 
exponential rotational operators up to first order we have
\begin{eqnarray}
\label{rotation2}
R_{_3}&=& 1-i\Delta\varphi \hat J_{_3},\nonumber\\
R_{_2}&=&1-i\Delta\vartheta \hat J_{_2},\nonumber\\
R_{_z}&=& 1-i\Delta\psi \hat J_{_z}=1-i\Delta\psi(\cos{\vartheta}\hat J_{_3}+
\sin{\vartheta}\hat J_{_1}).
\end{eqnarray}
In this linear order the rotation operators $R_i$  commute and we can arrange them in a new
set that performs small rotations about the axes 1,2 and 3:
\begin{eqnarray}
\label{rotation3}
\tilde R_{_3}&=& 1-i(\Delta\varphi +\Delta\psi \cos{\vartheta})\hat J_{_3},\nonumber\\
 R_{_2}&=&1-i\Delta\vartheta \hat J_{_2},\nonumber\\
\tilde R_{_1}&=& 1-i\Delta\psi \sin{\vartheta}\hat J_{_1}.
\end{eqnarray}
We begin with the calculation of the change of the expectation values 
of the components $\hat Q_{_{\pm 2}}$ with respect to the rotated state $\tilde R_{_3}|TAC \rangle$. 
Using  Eqs.~(\ref{rotation3}) we obtain
\begin{eqnarray}
\langle \tilde R^\dagger_3 \hat Q_{_{\pm 2}}\tilde R_3\rangle=\langle \hat Q_{_{\pm 2}}\rangle +
i (\Delta\varphi +\Delta\psi \cos{\vartheta})\langle[\hat J_{_3},\hat Q_{_{\pm 2}}]\rangle
=\langle \hat Q_{_{\pm 2}}\rangle \pm 2i (\Delta\varphi +\Delta\psi \cos{\vartheta})
\langle \hat Q_{_{\pm 2}}\rangle.
\end{eqnarray}
For the hermitian operators $\hat Q_{_{2\pm}}$ defined by Eqs.~(\ref{combin})  it follows 
\begin{eqnarray}
\Delta Q_{_{2-}}&=& 2 (\Delta\varphi +\Delta\psi \cos{\vartheta})
\langle \hat Q_{_{2+}}\rangle ,\label{dQ2-phi}\\
\langle \tilde R^\dagger_{_3} \hat Q_{_{2+}}\tilde R_{_3}\rangle&=&
\langle \hat Q_{_{2+}}\rangle\label{dQ2+phi},
\end{eqnarray}
where $\langle \hat Q_{_{2-}}\rangle=0$ was used.
Eq.(\ref{dQ2-phi}) shall be used below to determine $\Delta\varphi$ whereas 
the last relation (\ref{dQ2+phi}) expresses  the fact that the rotation 
$\tilde R_{_3}$ does not lead to a shape change $\Delta Q_{_{2+}}$. The same  is valid  for the other 
components 
\begin{eqnarray}
\langle \tilde R^\dagger_3 \hat Q_{_{\pm 1}}\tilde R_3\rangle&=&\langle \hat Q_{_{\pm 1}}\rangle +
i \Delta\varphi \langle[\hat J_{_3},\hat Q_{_{\pm 2}}]\rangle
=\langle \hat Q_{_{\pm 1}}\rangle \pm i (\Delta\varphi +\Delta\psi \cos{\vartheta})
\langle \hat Q_{_{\pm 1}}\rangle=0,\\
\langle \tilde R^\dagger_3 \hat Q_{_{0}}\tilde R_3\rangle&=&\langle \hat Q_{_{0}}\rangle,
\end{eqnarray}
where $\langle \hat Q_{_{1\pm}}\rangle=0$ was used.
Now we calculate the components of $\hat Q_{_{\pm 1}}$ for the rotated state $R_{_2}|TAC\rangle$.  We have 
\begin{eqnarray}
\langle R^\dagger_{_2} \hat Q_{_{\pm 1}}R_{_2}\rangle=
i \Delta\vartheta \langle[\hat J_{_2},\hat Q_{_{\pm 1}}]\rangle
= \pm \frac{\Delta\vartheta}{2}\,(2\langle \hat Q_{_{2}}\rangle
-\sqrt{6}\langle \hat Q_{_{0}}\rangle)=\Delta Q_{_{\pm 1}},
\end{eqnarray}
where we used $\hat J_{_2}=-i/2\,(\hat J_{_+}-\hat J_{_-})$ and $\langle \hat Q_{_{\pm 1}}\rangle
 =0$ . This gives for  the hermitian combination $\hat Q_{_{1-}}$  
\begin{eqnarray}
\Delta Q_{_{1-}}=\Delta\vartheta \,(\langle \hat Q_{_{2+}}\rangle
-\sqrt{3}\langle \hat Q_{_{0}}\rangle),
\end{eqnarray}
which yields expression for the amplitude $\Delta\vartheta$, Eq.(\ref{anglefluc2}). The
other components of the quadrupole tensor do not change because 
$\langle[\hat J_\pm,\hat Q_{\pm2}]\rangle$ and $\langle[\hat J_\pm,\hat Q_{0}]\rangle$ 
are proportional to $\langle\hat Q_{\pm1}\rangle=0$. 
Finally we calculate the expectation values of the quadrupole operators $\hat Q_{_{\pm 1}}$ 
for $\tilde R_{_3}|TAC\rangle $.
Using $\hat J_{_1}=1/2\,(\hat J_{_+}+\hat J_{_-})$ we obtain analogously
\begin{eqnarray}
  \langle \tilde R^\dagger_{1} \hat Q_{_{\pm 1}}\tilde R_{1}\rangle=
i \Delta\psi\,\langle[\sin{\vartheta} \hat J_{_1},
\hat Q_{_{\pm 1}}]\rangle
=i \frac{\Delta\psi}{2}\,\sin{\vartheta}\,(2\langle \hat Q_{_{2}}\rangle
+\sqrt{6}\langle \hat Q_{_{0}}\rangle).
\end{eqnarray}
 Considering the combination $\hat Q_{_{1+}}$
we get
\begin{eqnarray}
\Delta Q_{_{1+}}=\Delta\psi\sin{\vartheta}\, (\langle \hat Q_{_{2+}}\rangle
+\sqrt{3}\langle \hat Q_{_{0}}\rangle),
\end{eqnarray}
which gives the amplitude $\Delta\psi$ by Eq.(\ref{anglefluc3}). With help of Eq.(\ref{dQ2-phi}) the 
amplitude $\Delta\varphi$ given by Eq.(\ref{anglefluc1}) is found.

 }

\begin{thebibliography}{99} 
\bibitem{Fr01} 
  S.~Frauendorf, Rev. Mod. Phys. {\bf 73}, 463 (2001) 
\bibitem{DF00}  
  V.I.~Dimitrov, S.~Frauendorf and F.~D\"{o}nau, Phys. Rev. Lett {\bf 84}, 5732 (2000) 
\bibitem{OD04} 
  P.~Olbratowski, J.~Dobaczewski, J.~Dudek and W.~Plociennik,  
  Phys. Rev. Lett. {\bf 93}, 052501 (2004) 
\bibitem{FM97} 
  S.~Frauendorf and J.~Meng, Nucl. Phys. {\bf A617}, 131 (1997) 
\bibitem{SC02} 
  K.~Starosta {\it et al.}, Phys. Rev. C {\bf 65}, 044328 (2002) 
\bibitem{BV04} 
  S.~Brant, D.~Vretenar and A.~Ventura, Phys. Rev. C {\bf 69}, 017304 (2004) 
\bibitem{TA06}  
  D.~Tonev {\it et al.}, Phys. Rev. Lett. {\bf 96}, 052501 (2006) 
\bibitem{SK01}  
  K.~Starosta {\it et al.}, Phys. Rev. Lett {\bf 86}, 971 (2001) 
\bibitem{HB01} 
  A.A.~Hecht {\it et al.}, Phys. Rev. C {\bf 63}, 051302(R) (2001) 
\bibitem{RS80} 
  P.~Ring and P.~Schuck, {\em The Nuclear Many-Body Problem},  
  (Springer, New York, 1980)  
\bibitem{PH06} 
  C.M.~Petrache, G.B.~Hagemann, I.~Hamamoto and K.~Starosta,  
  Phys.~Rev.~Lett. {\bf 96}, 112502 (2006) 
\bibitem{GS06} 
  E.~Grodner {\it et al.}, Phys. Rev. Lett. {\bf 97}, 172501 (2006) 
\bibitem{MA07} 
  S.~Mukhopadhyay {\it et al.}, Phys. Rev. Lett. (2007) 
\bibitem{ZG03} 
  S.~Zhu {\it et al.}, Phys. Rev. Lett. {\bf 91}, 132501 (2003) 
\bibitem{Fr00} 
  S.~Frauendorf, Nucl. Phys. {\bf A677}, 115 (2000) 
\bibitem{BK68}
  M.~Barranger and K.~Kumar, Nuc.~Phys. {\bf A110},  490 (1968) 
\bibitem{Nilsson-Ragnarsson} 
S. G. Nilsson and I. Ragnarsson, Shapes and Shells in Nuclear
Physics (Cambridge University Press, Cambridge, 1995).
\bibitem{KN86}.
  J.~Kvasil and R.~Nazmitdinov, Sov. J. Part. Nucl. {\bf 17}, 265 (1986) 
\bibitem{KS03} 
  T.~Koike, K.~Starosta, C.J.~Chiara, D.B.~Fossan, and D.R.~LaFosse,  
  Phys. Rev. C {\bf 67}, 044319 (2003) 
\bibitem{HR01} 
  D.J.~Hartley {\it et al.}, Phys. Rev. C {\bf 64}, 031304(R) (2001) 
\bibitem{KS01} 
  T.~Koike, K.~Starosta, C.J.~Chiara, D.B.~Fossan, and D.R.~LaFosse,  
  Phys. Rev. C {\bf 63}, 061304(R) (2001) 
\bibitem{BB01} 
  R.A.~Bark {\it et al.}, Nucl. Phys. {\bf A 691}, 577 (2001) 
\bibitem{ADF11} 
  D.~Almehed, F. D\"onau,  and S.~Frauendorf, to be published
\bibitem{TA07} 
   D.~Tonev {\it et al.}, Phys. Rev. C {\bf 76}, 044313 (2007) 
  \bibitem{Negele82}
   J. W. ~Negele, Rev. Mod. Phys. {\bf 54}, 914 (1982)
\bibitem{SM96}
  Y.R.~Shimizu and M.~Matsuzaki, Nucl. Phys. {\bf A588}, 559 (1996)
\end{thebibliography}
\end{document}